\documentclass[longauth]{aa} 

\newcommand{\ltsim}{\protect\raisebox{-0.5ex}{$\:\stackrel{\textstyle <}
        {\sim}\:$}}

\usepackage{longtable}    
\usepackage{latexsym}
\usepackage{graphicx}
\usepackage{epsfig}
\usepackage{color}
\usepackage{lscape}
\usepackage{natbib}
\bibpunct{(}{)}{;}{a}{}{,} 

\begin{document}

\title{The Gaia-ESO Survey: Chromospheric Emission, Accretion Properties, and Rotation in $\gamma$~Velorum and Chamaeleon~I\thanks{Based on data 
products from observations made with ESO Telescopes at the La Silla Paranal Observatory under programme ID 188.B-3002. }\fnmsep\thanks{Figures 
\ref{fig:vsini_test}--\ref{fig:spectrum_accretor} and Tables \ref{Tab:gammavel}--\ref{tab:Macc_literature}
are only available in electronic form at {\tt http://www.aanda.org}.}}
   
\author{A. Frasca\inst{\ref{inst:oact}}	
\and K. Biazzo \inst{\ref{inst:oact}}		 
\and A.~C. Lanzafame\inst{\ref{inst:unict},\ref{inst:oact}}	
\and J.~M. Alcal\'a\inst{\ref{inst:oana}}	 
\and E. Brugaletta\inst{\ref{inst:unict},\ref{inst:oact}}	 
\and A. Klutsch\inst{\ref{inst:oact}}		 
\and B. Stelzer\inst{\ref{inst:oapa}}		 
\and G.~G. Sacco\inst{\ref{inst:oaar}}	 
\and L. Spina\inst{\ref{inst:oaar}}	 
\and R.~D. Jeffries\inst{\ref{inst:keele}}	 
\and D. Montes\inst{\ref{inst:compl}}		 
\and E. J. Alfaro\inst{\ref{inst:grana}}	
\and G. Barentsen\inst{\ref{inst:hert}}	 
\and R. Bonito\inst{\ref{inst:unipa},\ref{inst:oapa}}
\and J.~F. Gameiro\inst{\ref{inst:porto}}	
\and J. L\'opez-Santiago\inst{\ref{inst:madrid}}
\and G. Pace\inst{\ref{inst:porto}}
\and L. Pasquini\inst{\ref{inst:eso}}	
\and L. Prisinzano\inst{\ref{inst:oapa}} 
\and S.~G. Sousa\inst{\ref{inst:porto},\ref{inst:porto2}}	 
\and G. Gilmore\inst{\ref{inst:cambr}}
\and S. Randich\inst{\ref{inst:oaar}}
\and G. Micela\inst{\ref{inst:oapa}}
\and A. Bragaglia\inst{\ref{inst:oabo}}
\and E. Flaccomio \inst{\ref{inst:oapa}}
\and A. Bayo\inst{\ref{inst:maxpl},\ref{inst:valpa}}
\and M.~T. Costado\inst{\ref{inst:grana}}
\and E. Franciosini\inst{\ref{inst:oaar}}
\and V. Hill\inst{\ref{inst:oca}}
\and A. Hourihane\inst{\ref{inst:cambr}}	
\and P. Jofr\'e\inst{\ref{inst:cambr}}
\and C. Lardo\inst{\ref{inst:oabo}}
\and E. Maiorca\inst{\ref{inst:oaar}}
\and T. Masseron\inst{\ref{inst:cambr}}
\and L. Morbidelli\inst{\ref{inst:oaar}}
\and C.~C. Worley\inst{\ref{inst:cambr}}	
}

\offprints{A. Frasca}

\institute{INAF - Osservatorio Astrofisico di Catania, via S. Sofia 78, 95123, Catania, Italy \email{antonio.frasca@oact.inaf.it}\label{inst:oact}
\and Dipartimento di Fisica e Astronomia, Sezione Astrofisica, Universit\`a di Catania, via S. Sofia 78, 95123, Catania, Italy\label{inst:unict}
\and INAF - Osservatorio Astronomico di Capodimonte, via Moiariello 16, 80131, Naples, Italy\label{inst:oana}  
\and INAF - Osservatorio Astronomico di Palermo, Piazza del Parlamento 1, 90134, Palermo, Italy\label{inst:oapa}
\and INAF - Osservatorio Astrofisico di Arcetri, Largo E. Fermi 5, 50125, Firenze, Italy\label{inst:oaar}
\and Astrophysics Group, Keele University, Keele, Staffordshire ST5 5BG, United Kingdom\label{inst:keele}
\and Departamento de Astrof\'{i}sica y Ciencias de la Atm\'osfera, Universidad Complutense de Madrid, 28040 Madrid, Spain\label{inst:compl}
\and Instituto de Astrof\'{i}sica de Andaluc\'{i}a, CSIC, Apdo 3004, 18080, Granada, Spain\label{inst:grana}
\and School of Physics, Astronomy \& Mathematics, University of Hertfordshire, College Lane, Hatfield, Hertfordshire, AL10 9AB, United Kingdom\label{inst:hert}
\and Dipartimento di Fisica e Chimica, Universit\`a di Palermo, Piazza del Parlamento 1, 90134, Palermo, Italy\label{inst:unipa}
\and Centro de Astrof\'{i}sica, Universidade do Porto, Rua das Estrelas 4150-752, Porto, Portugal\label{inst:porto}
\and S. D. Astronomía y Geodesia, Facultad de Ciencias Matem\'aticas, Universidad Complutense de Madrid, 28040 Madrid, Spain\label{inst:madrid}
\and European Southern Observatory, Karl-Schwartzschild-Strasse 2, 85748, Garching bei M\"unchen, Germany\label{inst:eso}
\and Departamento de F\'isica e Astronomia, Faculdade de Ci\^encias, Universidade do Porto, Rua do Campo Alegre, 4169-007 Porto, Portugal\label{inst:porto2}
\and Institute of Astronomy, University of Cambridge, Madingley Road, Cambridge CB3 0HA, United Kingdom\label{inst:cambr}
\and INAF - Osservatorio Astronomico di Bologna, via Ranzani 1, 40127, Bologna, Italy\label{inst:oabo}
\and Max-Planck Institut f\"{u}r Astronomie, K\"{o}nigstuhl 17, 69117, Heidelberg, Germany\label{inst:maxpl}  
\and Instituto de F\'isica y Astronom\'ia, Facultad de Ciencias, Universidad de Valpara\'iso, Av. Gran Breta\~na 1111, Playa Ancha, Valpara\'iso, Chile\label{inst:valpa} 
\and Laboratoire Lagrange (UMR7293), Universit\'e de Nice Sophia Antipolis, CNRS, Observatoire de la C\^ote d'Azur, CS 34229, F-06304 Nice cedex 4, France\label{inst:oca}
}

\date{Received / accepted }

\abstract
{}
{
One of the scopes of the Gaia-ESO Survey (GES), which is conducted with FLAMES at the VLT, is the census and the characterization of the
low-mass members of very young clusters and associations. 	
We conduct a comparative study of the main properties of the sources belonging to $\gamma$~Velorum ($\gamma$~Vel) and Chamaeleon~I (Cha~I) 
young associations, focusing on their rotation, chromospheric radiative losses, and accretion. 
}
{
We use the fundamental parameters (effective temperature, surface gravity, lithium abundance, and radial velocity) delivered by the GES consortium in the first 
internal data release to select the members of $\gamma$~Vel and Cha~I among the UVES and GIRAFFE spectroscopic observations.
 A total of 140 $\gamma$~Vel members and 74 Cha~I members were studied.
The procedure adopted by the GES to derive stellar fundamental parameters provided also measures of the projected rotational velocity ($v\sin i$). 
We calculated stellar luminosities through spectral energy distributions, while stellar masses were derived by comparison with evolutionary tracks. 
The spectral subtraction of low-activity and slowly rotating templates, which are rotationally broadened to match the $v\sin i$ of the targets, enabled us 
to measure the equivalent widths (EWs) and the fluxes in the H$\alpha$ and H$\beta$ lines. 
The H$\alpha$ line was also used for identifying accreting objects, on the basis of its equivalent width and the width at the 10\% of the line peak ($10\%W$), 
and for evaluating the mass accretion rate ($\dot M_{\rm acc}$).
} 
{The distribution of $v\sin i$ for the members of $\gamma$~Vel displays a peak at about 10 km\,s$^{-1}$ with a tail toward faster rotators.
There is also some indication of a different $v\sin i$ distribution for the members of its two kinematical populations. 
Most of these stars have H$\alpha$ fluxes corresponding to a saturated activity regime.  We find a similar distribution, but with a narrower peak, for
Cha\,I. Only a handful of stars in $\gamma$~Vel display signatures of accretion, while many more accretors were detected in the younger Cha~I, 
where the highest H$\alpha$ fluxes are mostly due to accretion, rather than to chromospheric activity.  
Accreting and active stars occupy two different regions in a $T_{\rm eff}$--flux diagram and 
we propose a criterion for distinguishing them. We derive $\dot M_{\rm acc}$ in the ranges 
$10^{-11}$--$10^{-9} M_\odot$\,yr$^{-1}$ and $10^{-10}$--$10^{-7} M_\odot$\,yr$^{-1}$ for $\gamma$~Vel and Cha~I accretors, respectively. 
We find less scatter in the $\dot M_{\rm acc}-M_\star$ relation derived through the H$\alpha$ EWs, when compared to the H$\alpha$ 
$10\%W$ diagnostics, in agreement with other authors.}
{}
   
	   
\titlerunning{Chromospheric Emission, Accretion, and Rotation of $\gamma$~Vel and Cha~I}
\authorrunning{A. Frasca et al.}
\maketitle

\section{Introduction}
\label{sec:intro}

During the pre-main sequence (PMS) evolutionary phase, solar-like and low-mass stars undergo remarkable changes of their internal structure, radius,
temperature and rotation velocity. Moreover, several phenomena affect their atmospheric layers and circumstellar environments with noticeable effects on the 
observed spectra. 

The rotation velocity distribution of stars in young clusters and associations is a fundamental tool to understand the relative importance of
the processes that lead the stars to spin up during their early life (contraction and mass accretion) over those which tend to slow down them 
(magnetic braking and disk locking).
Disks appear to regulate the stellar rotation only for  about the first 5 Myr of their life or less, when they are very frequent and
detected at infrared wavelengths \citep[e.g.,][and references therein]{ladaetal2006,siciliaetal2006} and accretion signatures, like strong and broad emission lines, 
are seen in the stellar spectra. After 5 Myr the disks dissipate quickly \citep[e.g.,][]{haischetal2001,Hernandez2008} and the stars are free to spin up 
as they contract and approach the zero-age main-sequence (ZAMS). The disk locking effect has been invoked as the responsible for the bimodal distribution 
of rotation periods observed in very young clusters for stars with $M> 0.25 M_{\sun}$ with the slower rotators being  very often objects with infrared 
excess from circumstellar disks \citep[e.g.,][]{herbstetal2002,rebulletal2002}. The presence of both slow and fast rotators is still observed in older 
clusters and associations with ages from about 30 to 200 Myr \citep[see, e. g.,][and references therein]{messinaetal2003,meibometal2009,messinaetal2010} 
and predicted by the models of angular momentum evolution \citep[e.g.,][]{bouvieretal1997,spadaetal2011}.

The magnetic activity is closely related to the stellar evolution during the PMS and main-sequence (MS) stages, and the resulting changes in the internal 
structure and surface rotation rate. Indeed, the dynamo mechanism generating the magnetic fields depends on the stellar rotation, 
differential rotation, and sub-photospheric convection. For stars in the MS phase, the level of magnetic activity, as expressed by the average chromospheric 
emission (CE), has been shown to decay with age due to magnetic braking, 
since the pioneering work of \citet{Skumanich1972}, who proposed a simple power 
law of the form CE\,$\propto t^{-1/2}$. Further works based on stars belonging to clusters and moving groups of different ages have proposed different 
relations between CE and age 
\citep[see, e.g.,][]{Soderblometal1991,PacePasquini2004,MamajekHillenbrand2008}. Recent indications support the 
CE decline with age till about 2 Gyr and a nearly constant  behaviour  thereafter \citep[e.g.,][]{Pace2013}.
 The age-activity-rotation relation has been also investigated by means of the X-ray coronal emission \citep[e.g.,][]{Pizzolato2003,Preibisch2005}. 
However, due to their faintness, very low-mass stars in many open clusters (OCs) and associations were only observed recently. 

The picture is more complicated for stars in the PMS phase, when accretion of material 
from the circumstellar disk onto the central star occurs. In particular, mass accretion in the early PMS evolution
 is responsible for a significant fraction of the final stellar mass and the time dependence of the mass 
accretion rate is important to trace the disk evolution and its dissipation, contributing to the conditions for both 
stellar and planetary formation  (e.g., \citealt{hartmann1998}).	
This implies that, during the PMS evolutionary phases, mass accretion affects the spectral diagnostics of CE, 
and, at the same time, chromospheric activity can be a source of contamination in
the measurements of mass accretion rates. The effects of accretion and magnetic activity on the optical emission lines 
become comparable at the final stages of the PMS evolution and in very low-mass stars (\citealt{calvetetal2005,bayoetal2012,inglebyetal2013}, and references therein).   
As recently found by \citet{manaraetal2013} for young disk-less (Class\,III) stellar objects with spectral types 
from mid-K to late M, the  CE, if misinterpreted as an effect of accretion, would give rise to mass accretion rates
($\dot M_{\rm acc}$)  ranging from $\sim 6.3\times 10^{-10} M_\odot$\,yr$^{-1}$ for solar-mass young ($\sim 1$ Myr) stars to 
$\sim 2.5\times 10^{-12} M_\odot$\,yr$^{-1}$ for low-mass older ($\sim 10$ Myr) objects.
 These authors consider this as a ``noise'' that is introduced by the CE or, equivalently, as a threshold for the detection of accretion.

The Gaia-ESO Survey \citep[GES,][]{gilmoreetal2012,randich_gilmore2013}  offers the possibility to considerably extend the dataset of low-mass PMS 
stars with intermediate- and high-resolution spectra.  Indeed, it is observing with FLAMES@VLT a very large sample ($\sim10^5$) of stars, surveying also
more than 70 OCs and star forming regions (SFRs) of different ages. 
The large number of members of the  nearby SFRs and young OCs  surveyed by the 
GES, besides the chemical composition and kinematics, enables us to make a comparative study of their basic 
properties like rotation velocity, level of magnetic activity, and incidence of mass accretion, which depend on	
 stellar mass and  cluster age.

As suitable laboratories to study the evolution of these parameters during the first 10\,Myr, we present here the case of  $\gamma$~Velorum 
(hereafter $\gamma$~Vel) and Chamaeleon~I (hereafter Cha~I), which are the first two young clusters observed within 
the GES.

$\gamma$~Vel is a nearby ($\sim 350$ pc) PMS OC with an age of 5--10 Myr and low extinction ($A_{V}=0.131$ mag, \citealt{jeffriesetal2009}). 
Its members are distributed around the double-lined high-mass spectroscopic binary system $\gamma^2$~Vel (\citealt{pozzoetal2000}). It belongs to 
the Vela OB2 association ($\alpha \sim 8^{\rm h}$, $\delta \sim -47^{\degr}$), a group of $\sim 100$ early-type stars spread over an angular 
diameter of $\sim 10\deg$ (see \citealt{dezeeuwetal1999}).   
Using the {\it Spitzer} mid-infrared (MIR) data, \citet{Hernandez2008} found a low frequency of circumstellar disks around low-mass stars. 
Moreover, the IR flux excess in $\gamma$ Vel is lower than that found in stellar populations with a similar age.
They propose that the strong radiation field and winds from the components of the 
$\gamma^2$~Vel binary could be responsible for a relatively fast dissipation of the  circumstellar dust around the nearby stars.

The Cha~I dark cloud, at a distance of $160\pm15$ pc (\citealt{whittetetal1997}), 
is one of the three main clouds of the Chamaeleon complex ($\alpha \sim 12^{\rm h}$, $\delta \sim -78^{\degr}$). It extends over a few square 
degrees in the sky and its population consists of 237 known members, including sub-stellar objects (see \citealt{luhman2008} for a recent review). 
As Cha~I is younger than $\gamma$~Vel (age\,$\sim 2$~Myr, \citealt{luhman2008}), this age difference allows us to perform a comparative 
analysis in terms of stellar activity and accretion.

This paper is based on results obtained by the GES on these two clusters in preparation to the first advanced 
data product release\footnote{see {\tiny\tt http://www.eso.org/sci/observing/phase3/data\_releases.html}}. The GES analysis of spectra in the field of young open 
clusters is described in \citet{lanzafameetal2014}, while some aspects relevant to the study of chromospheric activity, accretion and rotation are 
described in more details here. Furthermore,  we present results based on an alternative approach for the analysis of accretion 
that makes use of the line luminosity and that will be introduced in future GES data releases.

In Sect.~\ref{sec:obs} we briefly describe the data used in this paper and member selection. In Sect.~\ref{sec:analysis_results} 
the analysis of the projected rotation velocity, the veiling, the spectral energy distribution, the Hertzsprung-Russell (HR) diagram, the H$\alpha$ and H$\beta$ line equivalent widths 
and fluxes, and the mass accretion rate are reported. The discussion of our results on rotation, chromospheric emission, and 
accretion diagnostics is given in Sect.~\ref{sec:discussion}, while the conclusions are drawn in Sect~\ref{sec:conclusions}.

\section{Data}	
\label{sec:obs}
Our analysis is based on the products of spectroscopy obtained during the first six months of observations which are internally 
released to the members of the GES consortium in the GESviDR1Final catalog\footnote{{\tt http://ges.roe.ac.uk/}}. 

The target selection has been done according to the GES guidelines for the cluster observations \citep[see,][]{bragagliaetal2014}. 
The observations were performed using the CD\#3 cross disperser ($R=47\,000$, $\lambda=4764$--6820 \AA) for UVES and 
the HR15N grating setting ($R=17\,000$, $\lambda=6445$--6815 \AA) for GIRAFFE. A brief observing log is given in Table~\ref{tab:logs}.
A total of 1242 targets were observed with GIRAFFE in the field of $\gamma$~Vel, while in Cha~I GIRAFFE spectra of 674 stars were secured.
Far fewer spectra were acquired with UVES (80 targets in $\gamma$~Vel and 48 in Cha~I).
A detailed description of the target selection and spectroscopic observations is given by \citet{jeffriesetal2014} for $\gamma$~Vel and 
 \citet{spinaetal2014b} for Cha~I.

\citet{Saccoetal2014} describe the reduction procedure for the UVES spectra, while for the GIRAFFE ones we refer the reader to \citet{jeffriesetal2014} 
and Lewis et al. (2014, in prep.).

\begin{table*}
\caption{Summary of the GES observations of $\gamma$~Vel and Cha~I.  }
\label{tab:logs}
\begin{center}
\begin{tabular}{lcc|rr|rr}
\hline
\hline
   &    &   &   &  &   & \\
   &    &   &   \multicolumn{2}{c|}{$\gamma$~Vel}    &   \multicolumn{2}{c}{Cha~I}\\
Instrument &   Range & Resolution & \# stars & \# members & \# stars & \# members\\ 
	   &   (\AA) & ($\lambda/\Delta\lambda$) &  &  &  & \\ 
\hline
   &    &   &   &  &   & \\
UVES	&  4764--6820 & 47\,000 & 80   & 8   & 48  & 15 \\
GIRAFFE &  6445--6815 & 17\,000 & 1242 & 132 & 647 & 59 \\
\hline
\end{tabular}
\end{center}
\end{table*}

The spectra observed in the $\gamma$~Vel and Cha~I fields, 
 which are publicly available\footnote{{\tt \scriptsize http://www.eso.org/sci/observing/phase3/data\_releases.html}},
have been analyzed by the GES working groups WG8 and WG12.
WG8 derives radial velocity ($RV$) and projected rotational velocity ($v\sin i$)  both for GIRAFFE \citep{jeffriesetal2014} and
UVES spectra \citep{Saccoetal2014}. WG12  is the working group responsible for the analysis of PMS clusters and delivers spectral type 
($SpT$), effective temperature ($T_{\rm eff}$), surface gravity ($\log g$), $v\sin i$ (derived with a different approach than WG8), iron abundance ([Fe/H]), microturbulence ($\xi$), veiling ($r$), lithium equivalent width at $\lambda$6707.8~\AA~($EW_{\rm Li}$), lithium abundance ($\log n_{\rm Li}$), H$\alpha$/H$\beta$ equivalent widths 
($EW_{\rm H\alpha}$/$EW_{\rm H\beta}$) and fluxes ($F_{\rm H\alpha}$/$F_{\rm H\beta}$), H$\alpha$ full-width at 10\% of peak height (10\%$W_{\rm H\alpha}$), 
mass accretion rate ($\dot M_{\rm acc}$), and other elemental abundances ([X/H]).

\subsection{Member selection}
\label{sec:member_selection}

For the $\gamma$~Vel cluster, WG8 produced reliable values of $RV$ and $v\sin i$ for most of the 1242  targets observed with GIRAFFE. The analysis performed 
by WG12, restricted to spectra with S/N\,$\ge$\,20, provided values of the main stellar parameters ($SpT$, $T_{\rm eff}$, $\log g$)  for 1078 stars.  
Among the 80 stars observed with UVES, the stellar parameters were determined for 68 stars, the remaining being spectroscopic binaries (six stars) or 
early-type and rapidly-rotating stars. 

For the 674 stars observed with GIRAFFE in the Cha~I field, WG12 released values of the main stellar parameters for 556 of them, while 42 out of the 48 UVES 
sources have atmospheric parameter entries in the GESviDR1Final catalogue. As for $\gamma$~Vel,  the fundamental parameters were not derived for
the double-lined spectroscopic binaries (SB2s), the early-type and rapidly-rotating stars and all the sources that have a spectrum with S/N$<$\,20.
 
\begin{figure*}[ht]  
\begin{center}
\includegraphics[width=8.8cm]{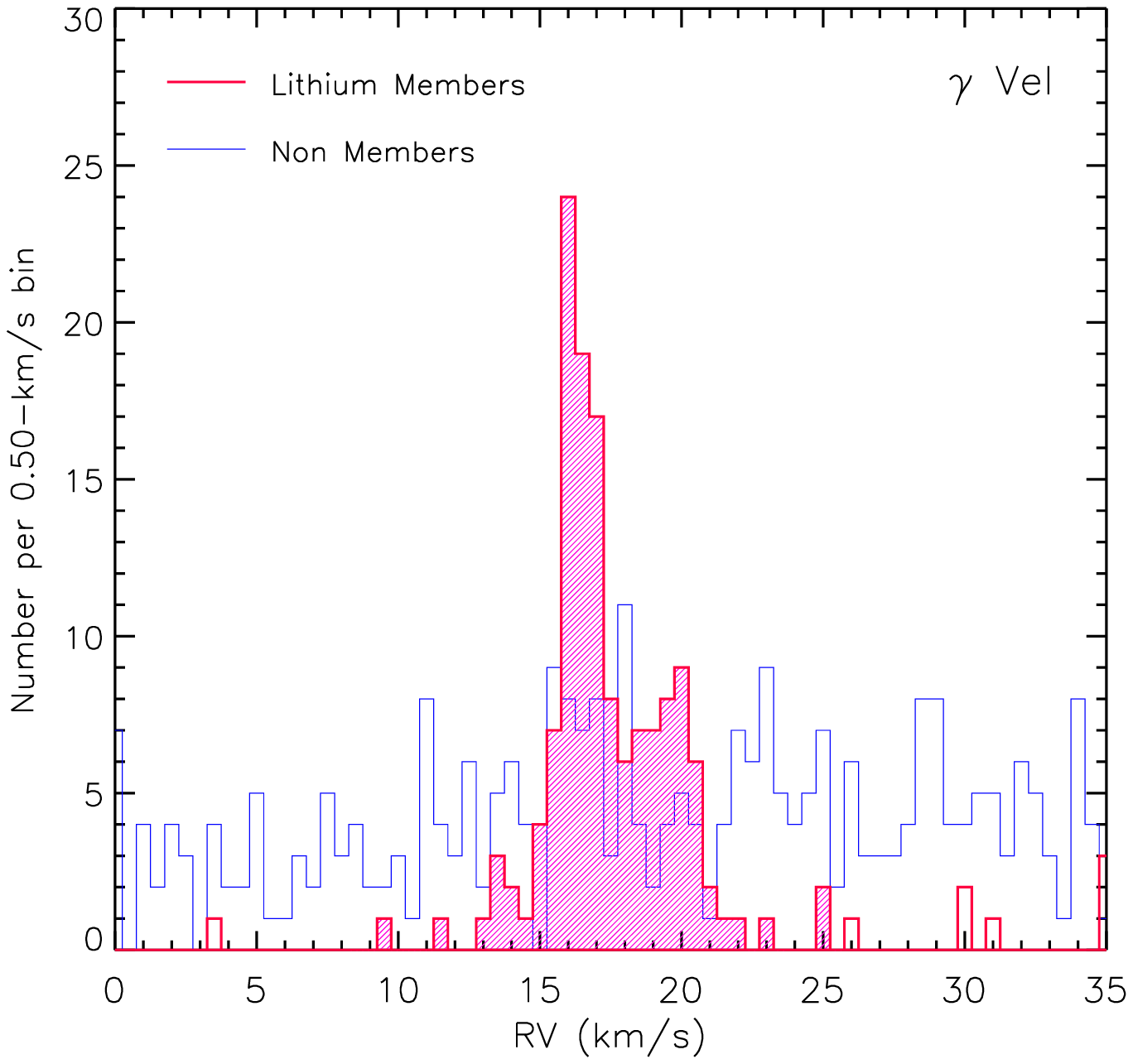}
\includegraphics[width=8.8cm]{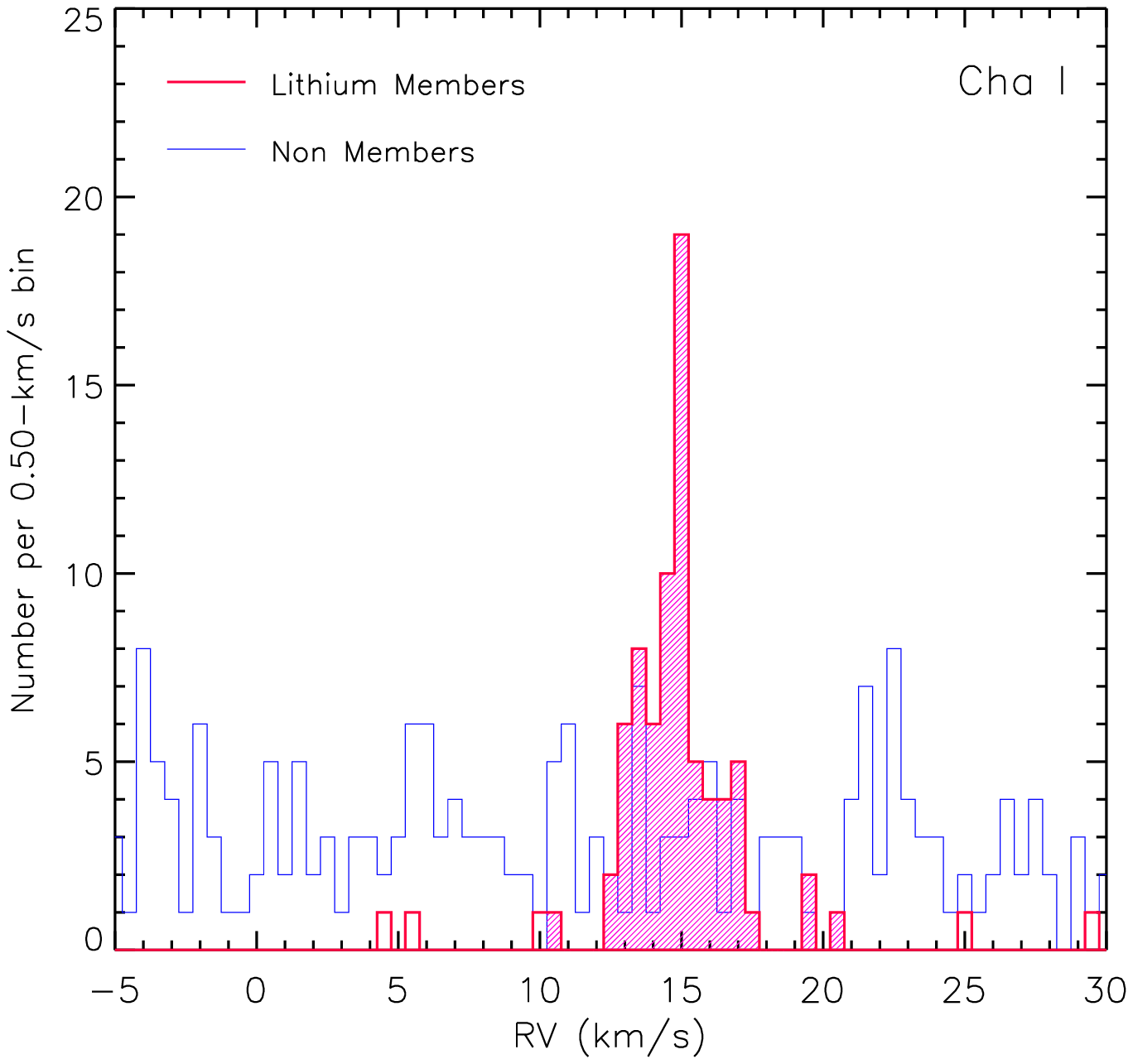}
\vspace{-.5cm}
\caption{Radial velocity distribution of $\gamma$~Vel ({\it left panel}) and Cha~I ({\it right panel}) stars. 
Thick (red) and thin (blue) 
lines represent the histograms of ``lithium members" and targeted non-members, respectively.	
 Lithium members fulfilling also the $RV$ criterion are represented by the red hatched area. Note in $\gamma$~Vel the double peak in the $RV$ distribution of members 
and the total absence of such peaks in that of non-members. }
\label{fig:vrad_distr_gammaCha_G}
 \end{center}
\end{figure*}

In the following, we use the membership criteria adopted by \cite{jeffriesetal2014} and \cite{spinaetal2014b} for $\gamma$~Vel and Cha~I, respectively. 
The selection performed by these authors was based on the strength of the lithium line at $\lambda$6707.8~\AA~(a reliable indicator of membership to 
 young clusters),  the surface gravity (to identify and discard lithium-rich giant contaminants in the field), and the position in the color-magnitude diagram 
(CMD, to recognize the cluster sequence). We refer to the objects pre-selected with these criteria as ``lithium members''. The final members are those
which fulfill an additional criterion based on their radial velocity: $8\leq RV \leq 26$\,km\,s$^{-1}$ and $10\leq RV \leq 21$\,km\,s$^{-1}$ for $\gamma$~Vel 
and Cha~I, respectively (see Fig.~\ref{fig:vrad_distr_gammaCha_G}).
The reader is referred to the aforementioned papers for a wide description of the membership analysis and the selection criteria. 
The only difference with respect to the aforementioned works is that we have slightly fewer targets, because we have restricted the analysis for deriving stellar parameters 
and chromospheric emission only to the spectra with a S/N larger than 20 per spectral point.
 The $RV$ distribution of the lithium members and targeted non-members of $\gamma$~Vel and Cha~I clusters, according to the above criteria is displayed in 
Fig.~\ref{fig:vrad_distr_gammaCha_G}. The finally selected members are represented by the hatched areas
in the histograms of Fig.~\ref{fig:vrad_distr_gammaCha_G}.

The parameters for the members of  $\gamma$~Vel and Cha~I clusters are reported in Tables~\ref{Tab:gammavel} and \ref{Tab:ChaI}, respectively.

In the end, our study is based on 132  members of $\gamma$~Vel cluster observed only with GIRAFFE (154 lithium members)
and on eight lithium members observed with UVES (six of them are also $RV$ members and two are also observed with GIRAFFE). 
For Cha~I, our analysis is based on 59 GIRAFFE and 15 UVES members.  We remark that the SB2s,  identified by means of the 
cross-correlation functions, are not included in our study. 	
However, the SB2s in our sample are rather few (28 in $\gamma$~Vel and 6 in Cha\,I) in comparison to the total number of targets
(both members and non-members) and most of them cannot be considered as candidate members on the base of lithium. Thus, we do not expect that 
their exclusion can have appreciably biased our sample. The same occurs for the rejection of the low S/N spectra. Compared to the 208 lithium members of 
$\gamma$~Vel reported by \citet{jeffriesetal2014}, we have 54 stars less, i.e. our sample is roughly 3/4 of that one. We are missing mostly some of the coolest stars, 
but this cut should have not biased the sample with respect to rotation velocity, H$\alpha$ flux, and accretion.

\section{Analysis}
\label{sec:analysis_results}

\subsection{Projected rotation velocity and veiling}
\label{sec:spec_analysis}

In the GES analysis of PMS clusters, $SpT$, $v\sin i$, and veiling are produced by one analysis node of WG12 that  makes use of ROTFIT, an 
IDL\footnote{IDL (Interactive Data Language) is a registered trademark of Exelis Visual Information Solutions.} code developed for deriving $SpT$, $T_{\rm eff}$, 
$\log g$, [Fe/H], $r$, and $v\sin i$ of the targets. 
This code	compares the target spectrum with a grid of templates composed of high-resolution ($R\simeq 42\,000$) spectra of 294 
slowly-rotating, low-activity stars retrieved from the ELODIE Archive \citep{moultakaetal2004}. The templates were brought to the GIRAFFE resolution, aligned in wavelength with the target spectrum by means of the cross-correlation, resampled on its spectral points 
and artificially broadened by convolution with a rotational profile of increasing $v \sin i$ until the minimum of $\chi^2$ is reached (see \citealt{Frasca2003, Frasca2006}).
The list of templates along with their spectral type and atmospheric parameters is reported in Table~\ref{Tab:Elodie_templ}.

\begin{figure*}  
\begin{center}
\includegraphics[width=8.5cm]{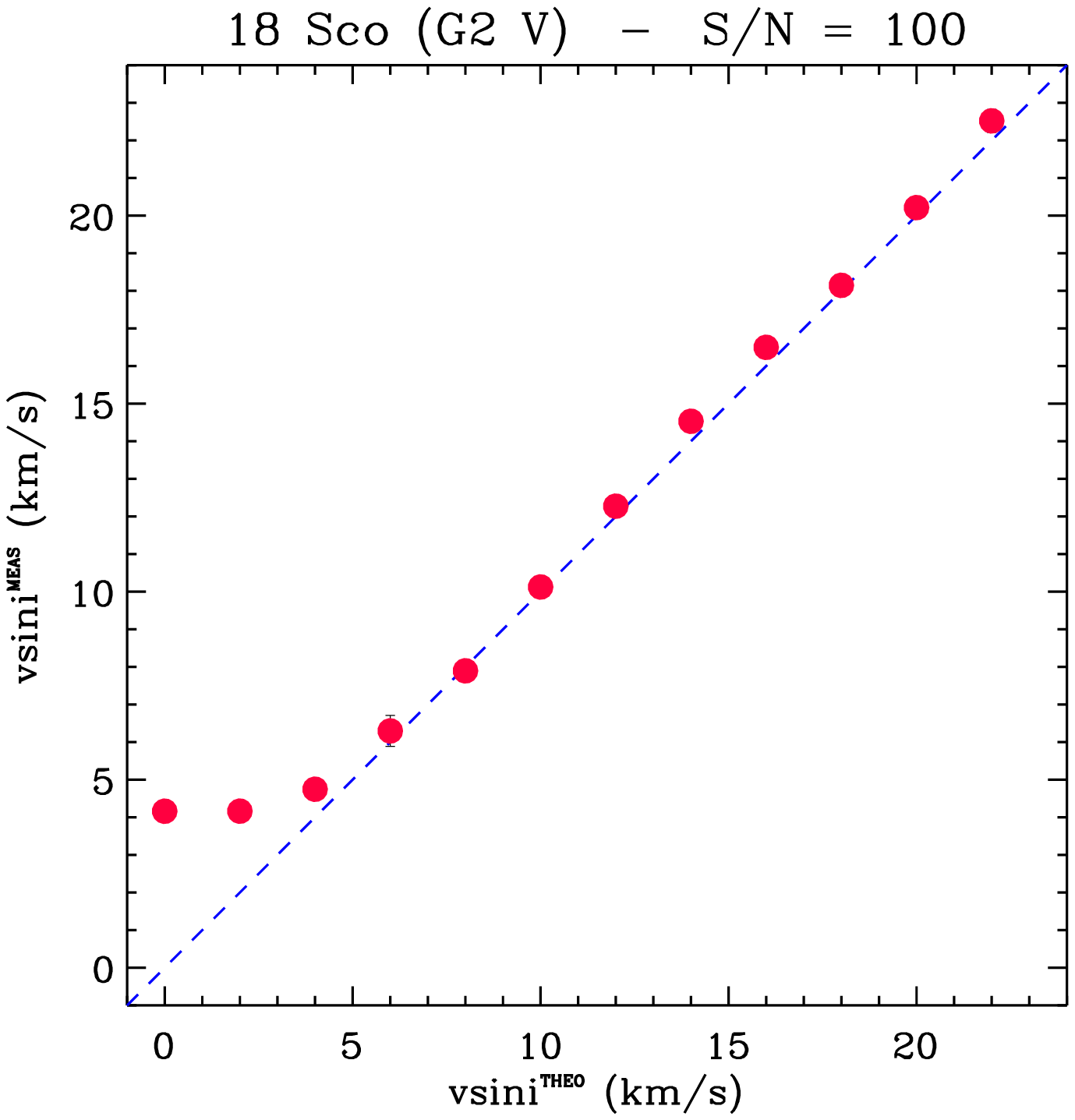}
\includegraphics[width=8.5cm]{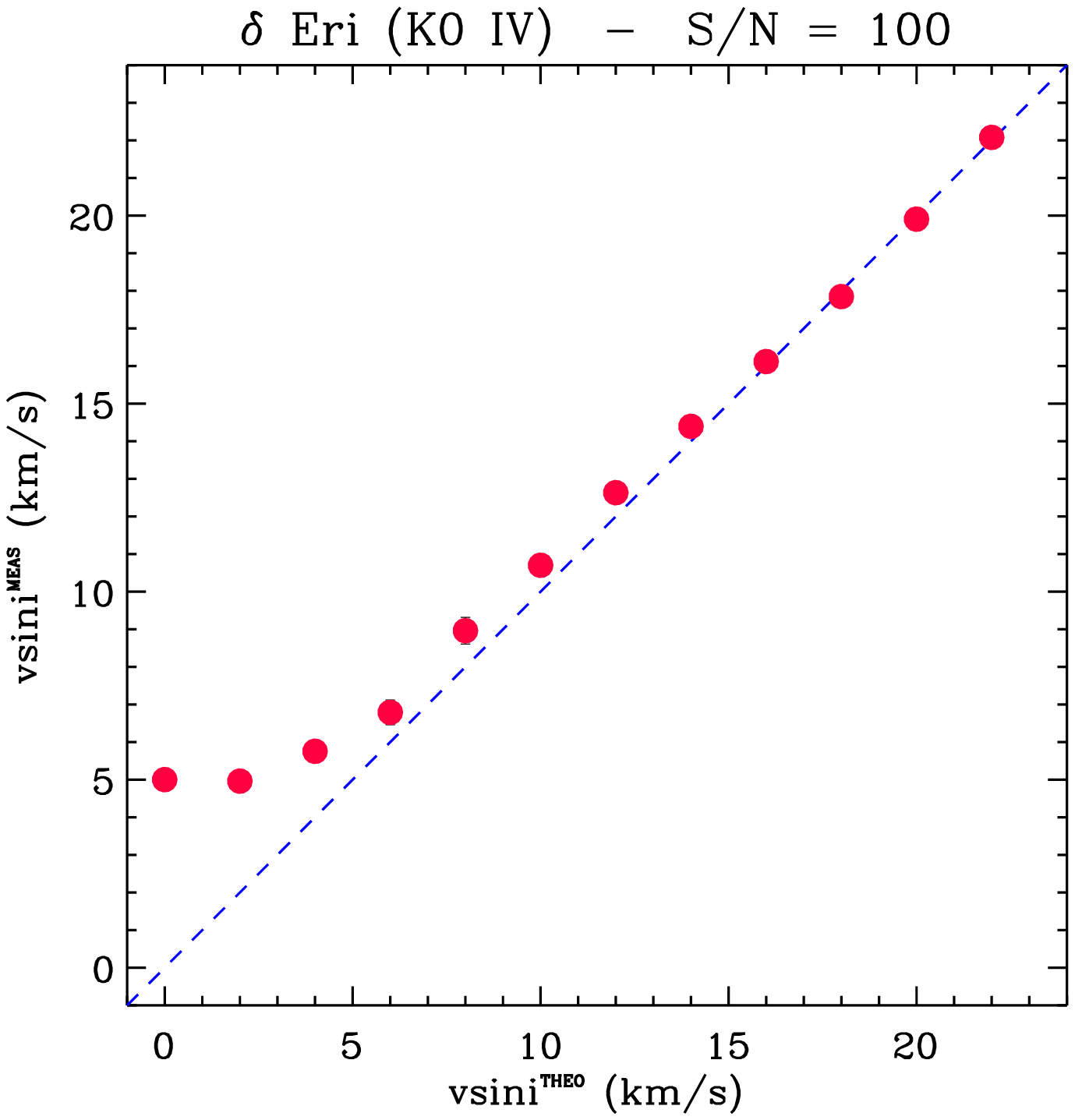}
\includegraphics[width=8.5cm]{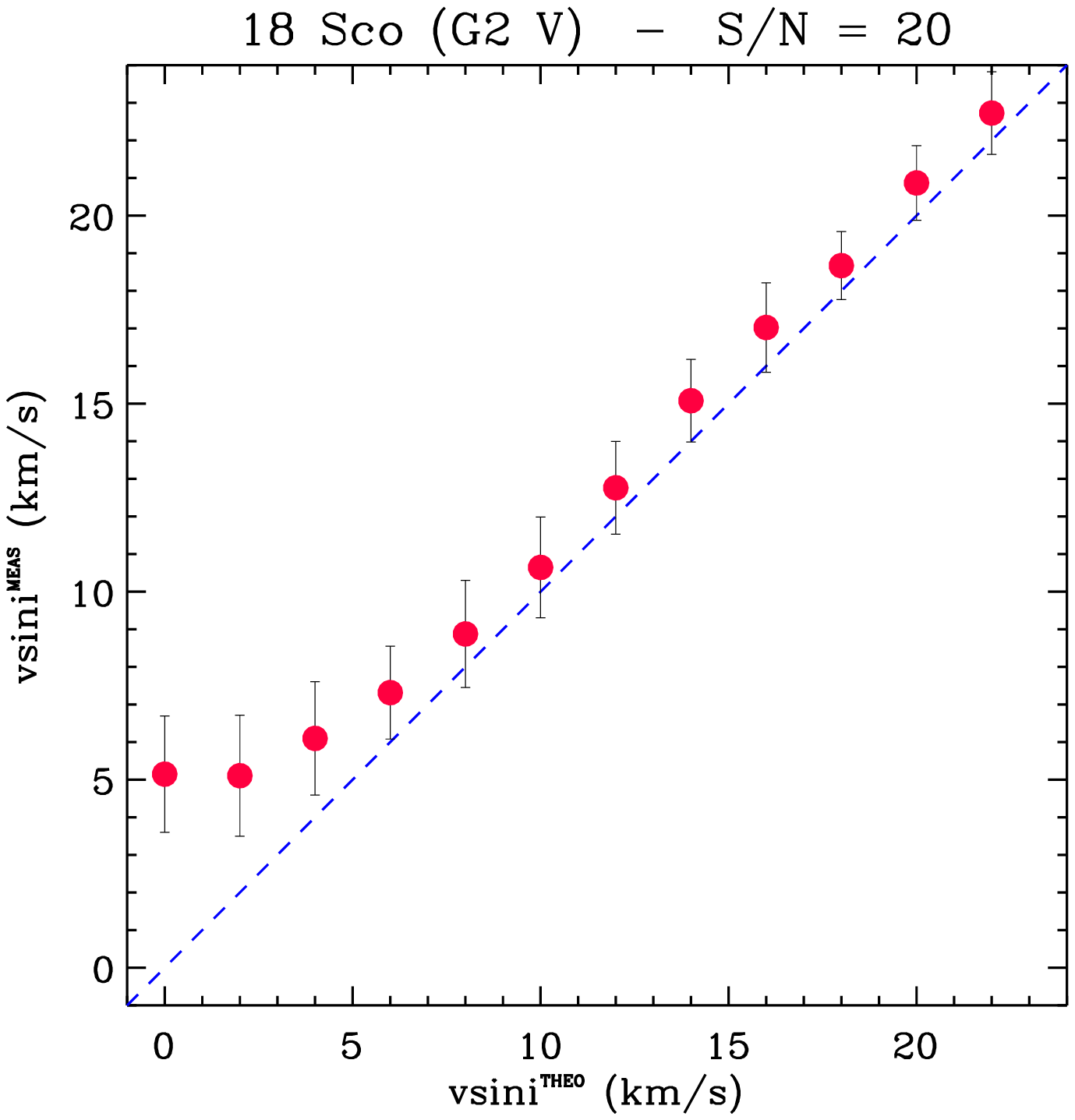}
\includegraphics[width=8.5cm]{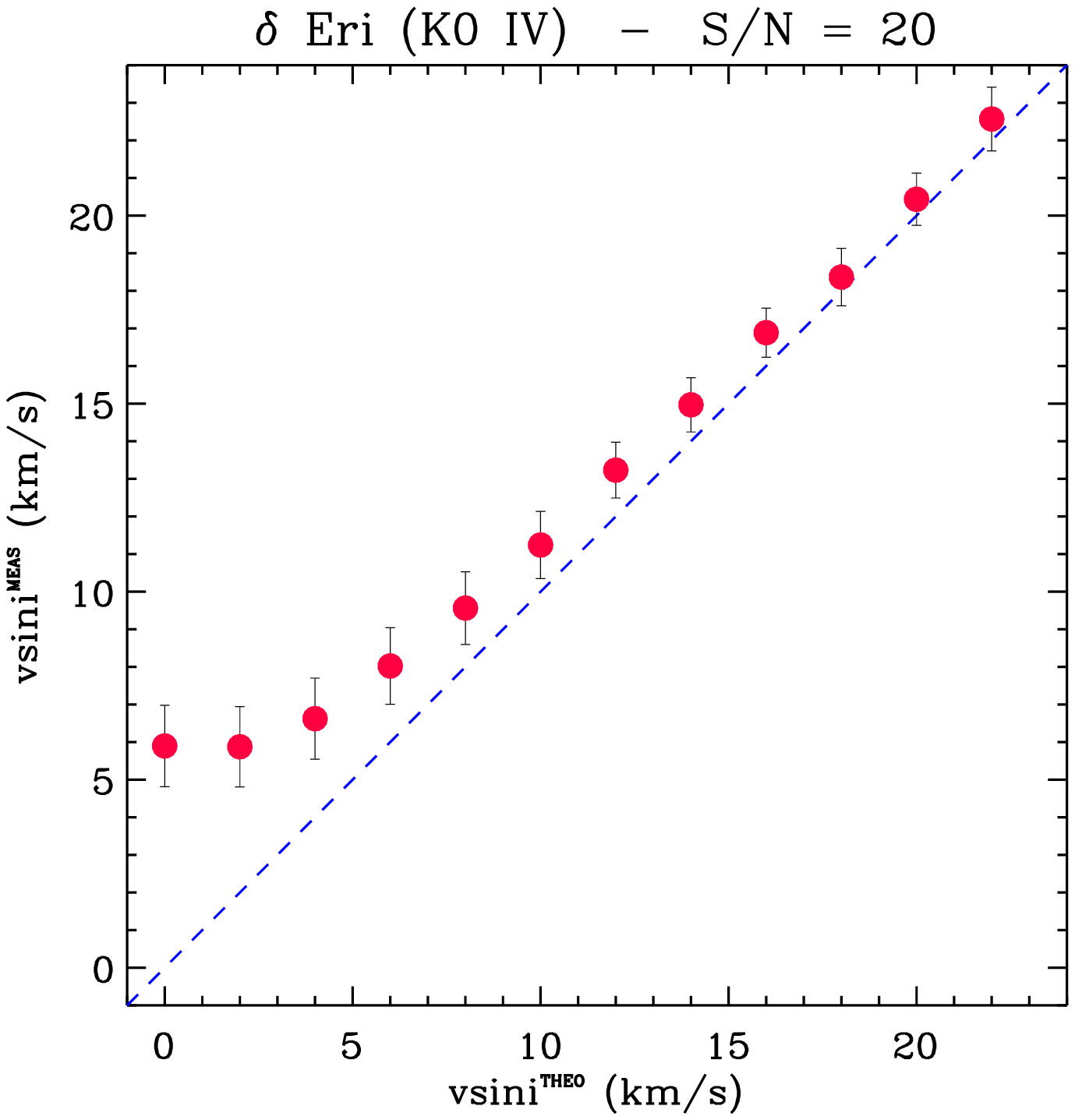}
\caption{Results of the Monte Carlo simulations on $v\sin i$ made with GIRAFFE spectra of two slowly-rotating stars for S/N=100 ({\it upper panels}) and S/N=20 
({\it lower panels}). The average $v\sin i$ measured with our procedure (dots) are plotted against the ``theoretical'' $v\sin i$ to which the spectra have been broadened. 
The one-to-one relation is plotted with a dotted line.}
\label{fig:vsini_test}
 \end{center}
\end{figure*}

To verify the ability of the procedure to derive the $v\sin i$ and to check the minimum detectable value with GIRAFFE spectra, we ran Monte Carlo 
simulations with two slowly-rotating stars, namely 18~Sco (G2\,V) and $\delta$~Eri (K0\,IV). The GIRAFFE spectra of these stars were artificially broadened by 
convolution with a rotation  profile of increasing $v\sin i$  (in steps of 2 km\,s$^{-1}$) and a random noise corresponding to a signal-to-noise ratio S/N=20 and S/N=100 was added. We made 100 simulations per each $v\sin i$ and S/N running ROTFIT on every simulated spectrum.
After the first nearly flat part where the $v\sin i$ is unresolved, the linear trend between measured and ``theoretical'' $v\sin i$ starts at 
6--8 km\,s$^{-1}$  (Fig.~\ref{fig:vsini_test}).  
We thus consider all the $v\sin i$ values lower than 7\,km\,s$^{-1}$ as upper limits.

\begin{figure}  
\begin{center}
\includegraphics[width=9.cm]{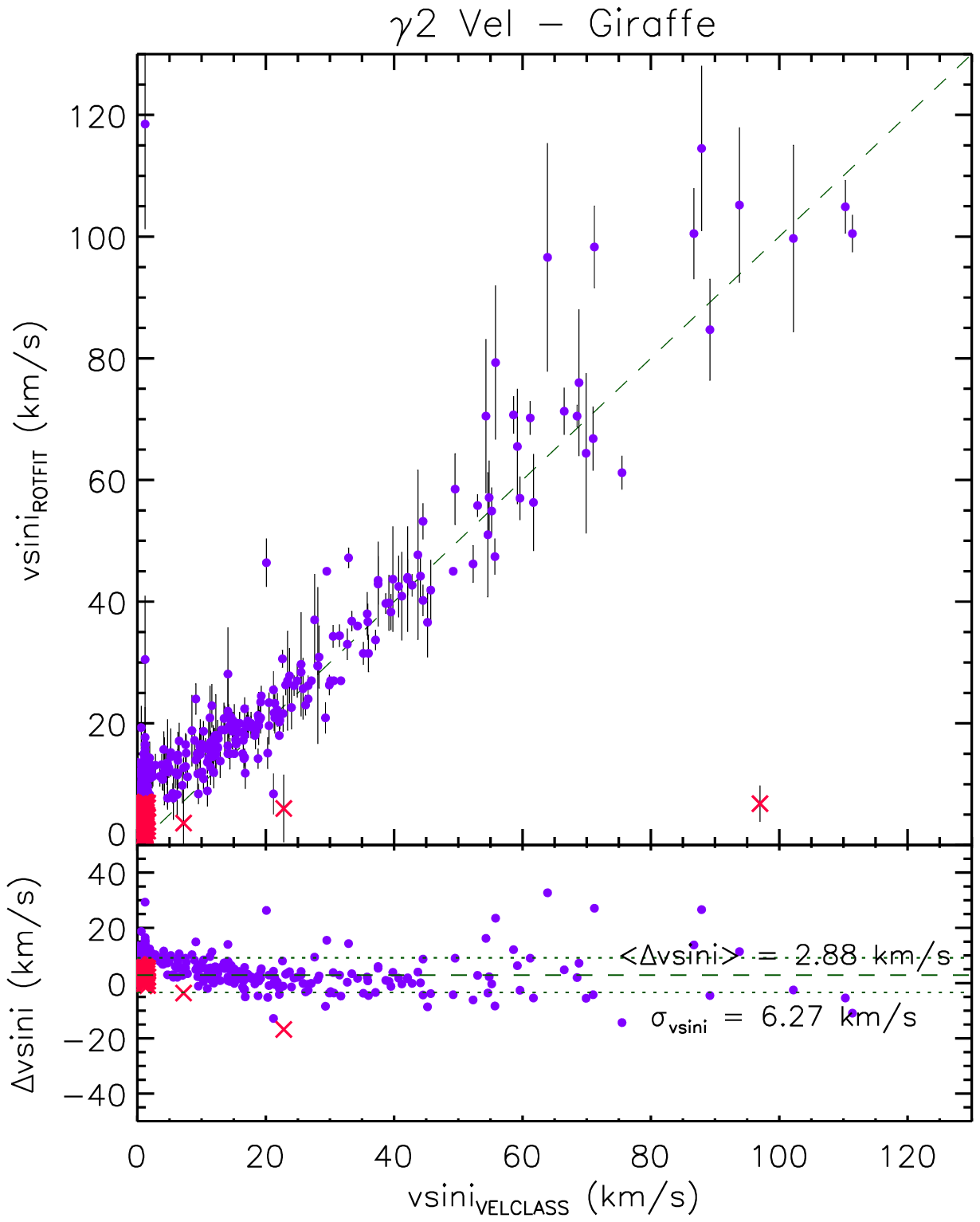}
\caption{Comparison between the $v\sin i$ measured by the VELCLASS (WG8) and ROTFIT (WG12) procedures for the stars in the $\gamma$\,Vel field.
 The ROTFIT's $v\sin i$ values lower than 7 km\,s$^{-1}$ are denoted by red crosses.}
\label{fig:vsini_comp}
 \end{center}
\end{figure}

As mentioned above, the $v\sin i$ for the GIRAFFE spectra is also measured, with a different procedure by WG8 along with the $RV$ determination
and the data are stored in the VELCLASS fits extension of the reduced spectra. The results of both procedures are compared for the stars in the $\gamma$\,Vel 
field in Fig.\,\ref{fig:vsini_comp}. The overall agreement between the two sets of values is apparent. 
However, in this study we use the $v\sin i$ values released by WG12.

The code ROTFIT is also able to evaluate the veiling of the spectra. Since this greatly increases the computing time, 
we have left $r$ free to vary in the code only when a veiling can be expected, i.e. for objects with a likely 
accretion.  
In the WG12, the ``accretor candidates'' are selected as those stars with $10\%W_{\rm H\alpha} \ge 270$ km\,s$^{-1}$ 
\citep[][]{whitebasri2003}. However, for these two young clusters, we preferred to use less restrictive criteria to check whether a significant 
veiling can be found by the code also for objects just under the above cutoff.
Thus, we ran the code with the veiling option enabled for all objects with $10\%W_{\rm H\alpha} \ge 200$ km\,s$^{-1}$.
We found a handful of objects with $200 < 10\%W_{\rm H\alpha} < 270$ km\,s$^{-1}$ and a non-zero veiling, all of which with  
$r < 0.25$, i.e. likely not significant.
The uncertainty of veiling determinations is in the range 20--50\,\% whenever $r > 0.25$ (see 
Tables~\ref{Tab:gammavel} and \ref{Tab:ChaI}).

\begin{figure}  
\begin{center}
\includegraphics[width=8.8cm]{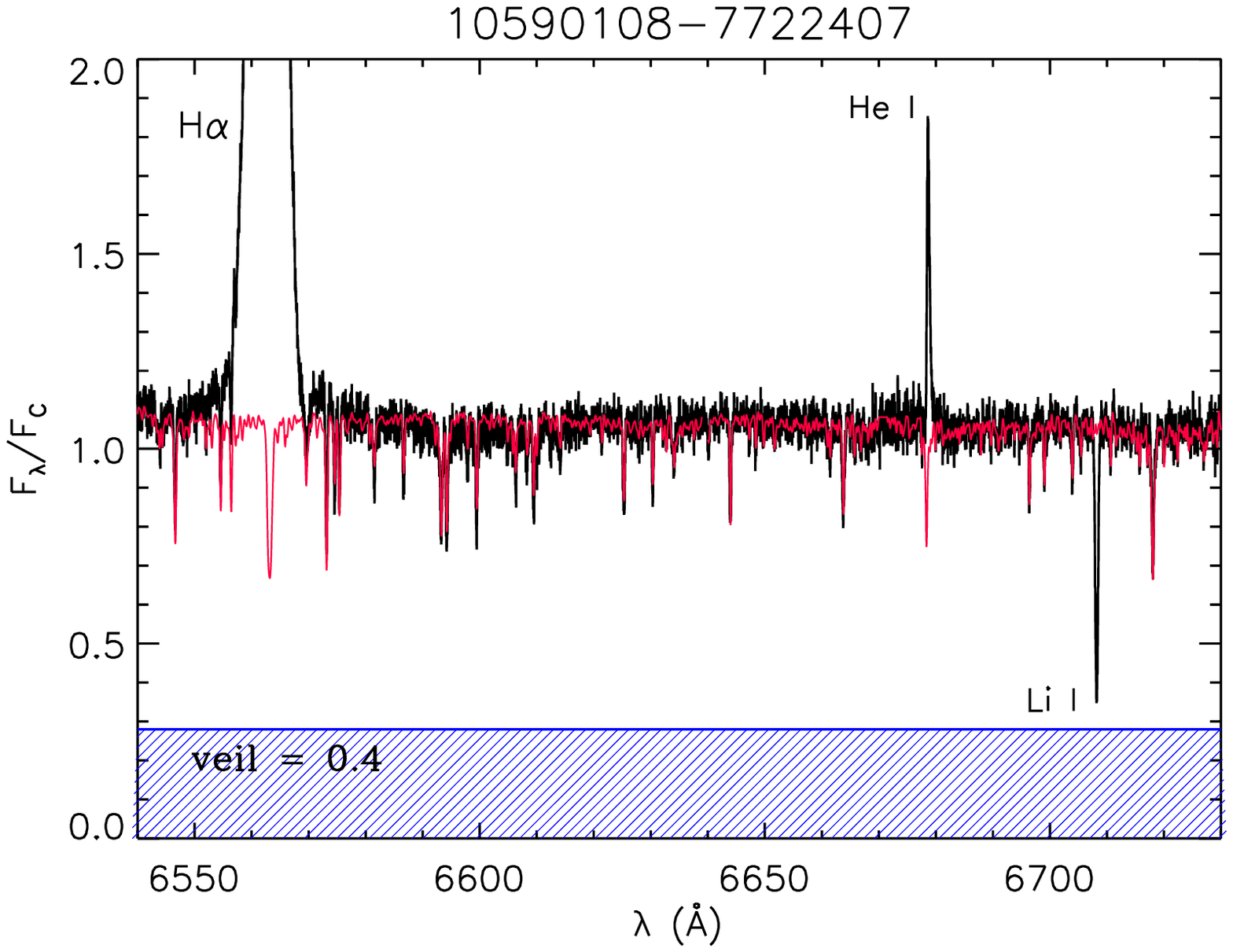}
\caption{UVES spectrum of an accreting star in Cha~I  (thick black line) with  overplotted the rotationally-broadened and veiled best template 
(thin red line). A wavelength independent veiling of 0.4 (hatched area) was found by ROTFIT.}
\label{fig:spectrum_accretor}
 \end{center}
\end{figure}

When searching for the best templates that reproduce the veiled stars, we considered the following equation:
\begin{equation}
\label{eq:veiling}
\biggl(\frac{F_\lambda}{F_C}\biggr)_{r}=\frac{\frac{F_\lambda}{F_C} + r}{1 + r}\,,
\end{equation}
where $F_\lambda$ and $F_C$ represent the line and continuum fluxes, respectively.
Moreover, $r$ was let free to vary to find the minimum $\chi^2$, assuming it constant over a limited wavelength range (which is 100 \AA\ for 
the 18 UVES spectral segments independently analyzed and about 300 \AA\ for the GIRAFFE spectra).
In Fig.~\ref{fig:spectrum_accretor} we show an example of an accreting star in Cha~I with $r=0.4$, as found by ROTFIT.

\begin{figure}[ht]  
\begin{center}
\includegraphics[width=8.7cm]{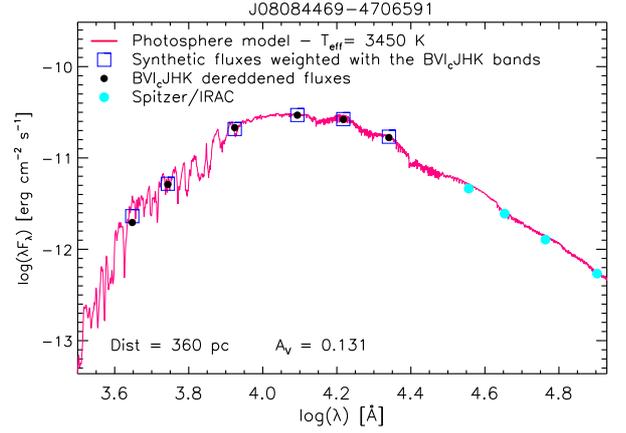}	
\includegraphics[width=8.7cm]{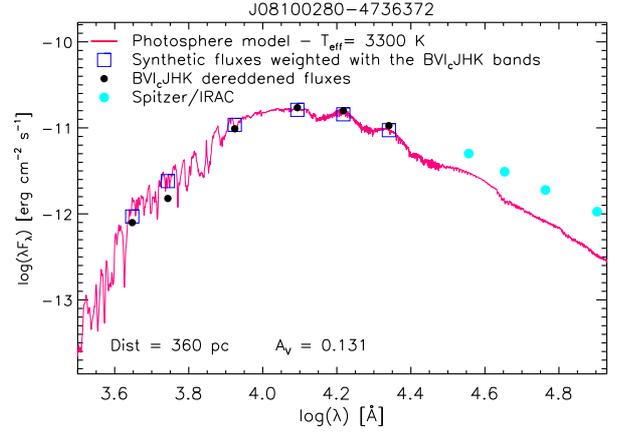}	
\includegraphics[width=8.7cm]{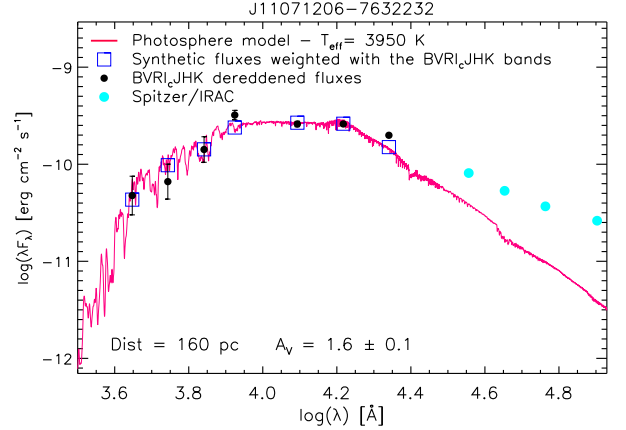}
\caption{Spectral energy distributions (dots) of two members of $\gamma$Vel cluster ({\it upper and middle panels}) and one member of the
young Cha~I association ({\it lower panel}). In each panel, the best fitting low-resolution NextGen spectrum \citep{hauschildtetal1999} is displayed by a continuous line. 
The SEDs of the two accretors ({\it middle and lower panels}) display a MIR excess, typical of Class\,II sources.
 }
\label{fig:SED}
 \end{center}
\end{figure}

We want to point out that the veiling is better and safer determined from the UVES spectra than from the GIRAFFE ones because the former 
have a much wider spectral coverage and include several strong lines suitable for the measurement of this parameter. 
This is testified by the internal agreement between values of veiling derived from adjacent segments (see also \citealt{biazzoetal2014}). 
In the case of HR15N GIRAFFE spectra, we can obtain only a rather rough estimate of veiling.

\subsection{Spectral Energy Distribution}
\label{sec:sed}

To obtain the stellar bolometric luminosities of all analyzed members of $\gamma$\,Vel and Cha~I, we constructed the spectral energy distribution 
(SED) of the targets using the optical and near-infrared (NIR) photometric data available in the literature. 

For the $\gamma$\,Vel stars, we combined optical $BVI_{\rm C}$ \citep{jeffriesetal2009} and 2MASS $JHK_{\rm s}$ (\citealt{skrutskieetal2006}) photometry. 
Moreover, $Spitzer$ mid-infrared (MIR) data from \citet{Hernandez2008} were also available for about 79\,\% of the sources. 
For the objects in Cha~I, we used $BVR$ photometry from the NOMAD 
catalog (\citealt{zachariasetal2004}) and Cousins $I_{\rm C}$ magnitudes from the DENIS database that were combined with 2MASS $JHK_{\rm s}$ and $Spitzer$
data \citep{luhmanetal2008}.
 
We then adopted the grid of NextGen low-resolution synthetic spectra, with $\log g$ in the range 3.5--5.0 and solar metallicity by \citet{hauschildtetal1999}, to 
fit the optical-NIR portion (from $B$ to $J$ band) of the SEDs,  similarly to what done by \citet{frascaetal2009} for stars in the Orion Nebula Cluster.

For the stars in the $\gamma$\,Vel cluster, we adopted the distance of 360\,pc and the extinction $A_V=0.131$\,mag found by \citet{jeffriesetal2009} and fixed 
the effective temperatures of the targets to the values found by the spectral analysis of WG12 and delivered in the first internal data release (iDR1).
We let the stellar radius ($R_\star$) vary until a minimum $\chi^2$ was reached. 
The stellar luminosity was then obtained by integrating the best-fit model spectrum.
We found a poor SED fitting only for three members of the cluster, namely J08101877$-$4714065, J08114456$-$4657516, and J08110328$-$4716357.   
This was likely due to a bad $T_{\rm eff}$ determination. For these stars we have used instead the photometric temperatures that are derived as described in 
\citet[][]{lanzafameetal2014}.  
For the members of Cha~I, which are scattered in a wide sky region with dense molecular clouds, we made the fit of the SEDs with the extinction parameter free to vary. 
The SEDs of two members of $\gamma$\,Vel, with and without MIR excess, and one of Cha~I are shown, as an example, in Fig.~\ref{fig:SED}.  

To our knowledge, no spectroscopic determination of effective temperaure from spectroscopy is available in the literature for the members of  $\gamma$\,Vel, 
while, for several objects in Cha~I, \citet{luhman2007} reported $SpT$, along with the corresponding $T_{\rm eff}$ values, derived from low-resolution spectroscopy. 
The comparison between our $T_{\rm eff}$ values and  Luhman's ones  shows a good agreement 
in the low temperature domain ($T_{\rm eff} <$3800--4000\,K), while a systematic difference appears for warmer stars in the sense that Luhman's values 
are lower than our ones by 200--400\,K (up to 800\,K in the worst case).  We think that the different spectral range and the lower resolution of Luhman's spectra can
be responsible for such a discrepancy. Moreover, some scatter could be also introduced by the binarity of a few sources \citep[e.g.,][]{Nguyen2012,daemgenetal2013}.
The bolometric luminosities, compared to the values reported by \citet{luhman2007}, do not show any relevant offset ($\approx -13$\%), but the rms deviation is rather large ($\approx66$\%). 

\subsection{HR diagram}
\label{sec:HRD}

\begin{figure}  
\begin{center}
\includegraphics[width=8.8cm]{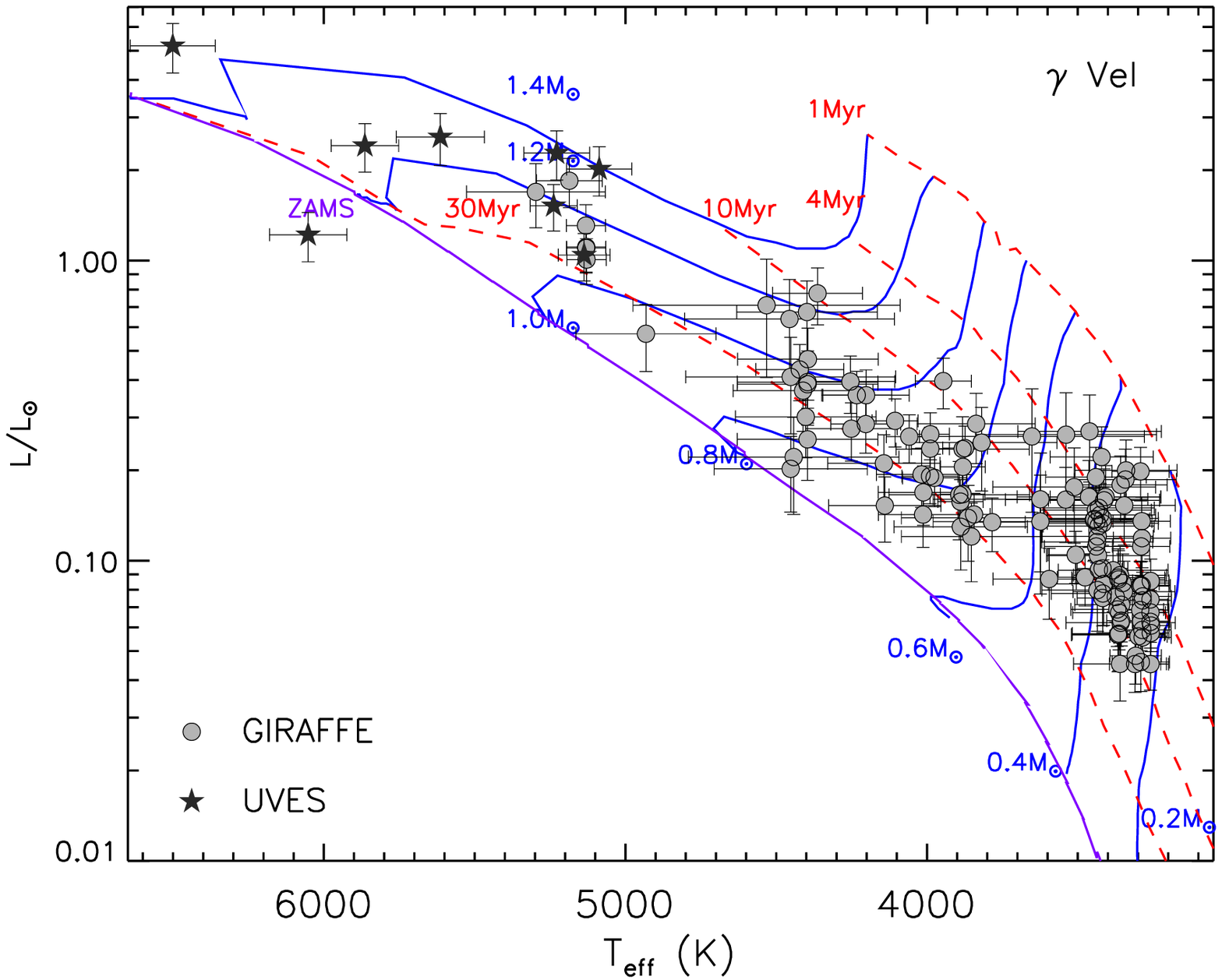}
\includegraphics[width=8.8cm]{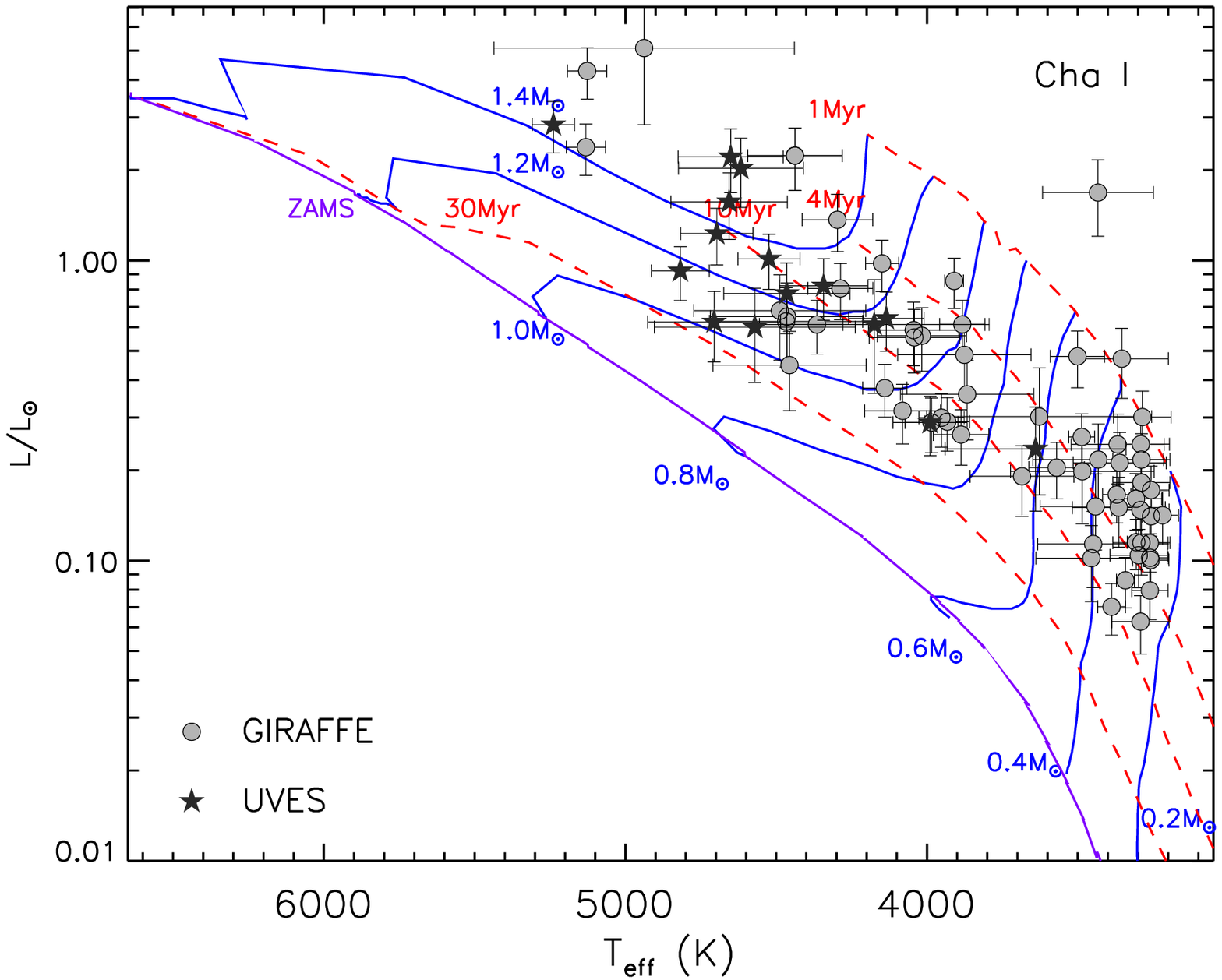}
\caption{HR diagram of $\gamma$~Vel ({\it upper panel}) and Cha~I ({\it lower panel}) members for both UVES and GIRAFFE data. The evolutionary 
tracks of \citet{baraffeetal1998} are shown by solid lines with the labels representing their masses. Similarly, the isochrones (from 1 to 30 Myr) 
by the same authors are shown with dashed lines. The ZAMS position is also represented by a solid line.}
\label{fig:HR_diagram}
 \end{center}
\end{figure}

In Figure~\ref{fig:HR_diagram}, we report the position of our targets in the HR diagram, along with the PMS evolutionary tracks 
and isochrones calculated by \citet{baraffeetal1998}. The effective temperatures are those from the WG12 analysis delivered in the iDR1, while the stellar 
luminosities are derived from the SED analysis illustrated in Sect.~\ref{sec:sed}. Most of the $\gamma$~Vel stars are located between the isochrones at 4 
and 30 Myr, while the Cha~I members lie higher in the diagram, as expected according to their younger age. 

We used the HR diagram and the evolutionary tracks for estimating the masses of the targets by minimizing the quantity:
\begin{equation}
\label{eq:HR_mass}
\chi^2 = \frac{(T_{\rm eff}-T_{\rm mod})^{2}}{\sigma_{T_{\rm eff}}^2} +\frac{(L-L_{\rm mod})^{2}}{\sigma_{L}^2} \,,
\end{equation}
where $T_{\rm eff}$ and $\sigma_{T_{\rm eff}}$ are the stellar effective temperature and its error, respectively, and $T_{\rm mod}$ is the temperature 
of the nearest evolutionary track. Analogously, $L$ and $\sigma_{L}$ are the stellar bolometric luminosity and its error, respectively, while $L_{\rm mod}$ is 
the luminosity of the nearest track.

 The masses, which are also reported in Tables~\ref{Tab:gammavel} and \ref{Tab:ChaI},  are used in Sect.~\ref{sec:accretion} to evaluate the 
mass accretion rate.

\subsection{Equivalent width and flux of the H$\alpha$ and H$\beta$ lines}
\label{sec:halpha_hbeta_flux}

\begin{figure}  
\begin{center}
\includegraphics[width=8.8cm]{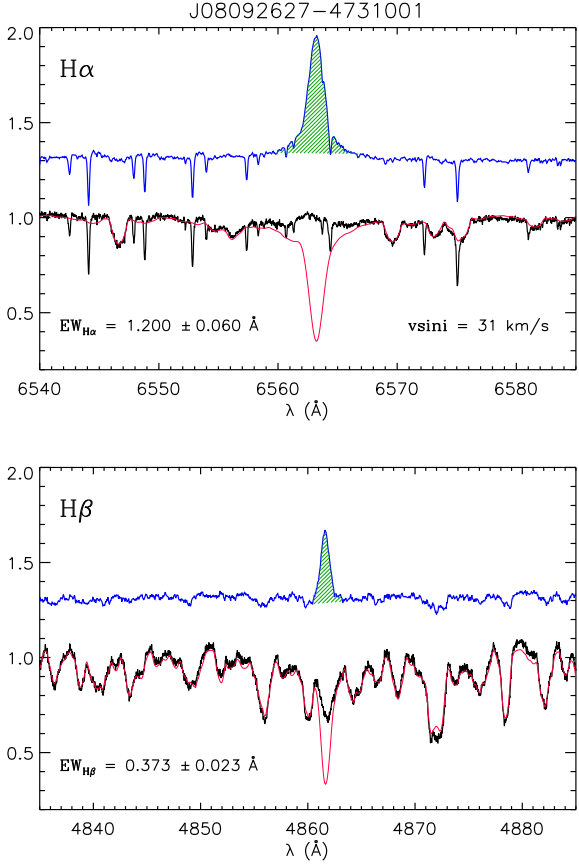}
\vspace{-1cm}
\caption{Example of the spectral subtraction method with UVES spectra in the H$\alpha$ ({\it upper panel}) and H$\beta$ ({\it lower panel}) regions for 
an active star in $\gamma$~Vel. 
The target spectrum is represented by a black solid line, while the best fitting reference spectrum of a low-activity star artificially broadened at the 
$v\sin i$ of the target is overplotted with a thin red line. In both panels, the difference spectrum (blue line) is displayed shifted upwards by 1.3 for clarity. 
The residual H$\alpha$ and  H$\beta$ profiles integrated over wavelength (hatched green areas) provide the net equivalent widths ($EW_{\rm H\alpha}$
and $EW_{\rm H\beta}$).
The narrow absorption features visible in the upper panel are telluric water vapor lines.
}
\label{fig:spectral_subtraction_U}
 \end{center}
\end{figure}

The most useful indicator of chromospheric activity in the HR15N GIRAFFE setup is the H$\alpha$ line, while the UVES spectra include 
also, among other diagnostics, the H$\beta$ line. Unlike the chromospheric and transition region lines at ultraviolet wavelengths, the contribution
of the photospheric flux in these optical lines is very important and must be removed to isolate the pure chromospheric emission that often 
is only filling-in the line cores. Thus, for the WG12 analysis, we have calculated EWs and fluxes by using
the spectral subtraction method (see, e.g., \citealt{frascacatalano1994, montesetal1995}, and references therein) to remove the photospheric flux and emphasize the 
chromospheric emission in the line core (see Fig.~\ref{fig:spectral_subtraction_U}).
Thanks to this procedure, the net equivalent width of the H$\alpha$ and H$\beta$ lines ($EW_{\rm H\alpha}$, $EW_{\rm H\beta}$) were derived 
(see Tables~\ref{Tab:gammavel}, \ref{Tab:ChaI}, and \ref{Tab:Hbeta}).  
In the example shown in Fig.~\ref{fig:spectral_subtraction_U}, the H$\alpha$  line is totally filled-in by emission and its core is 
just reaching the continuum level, while the H$\beta$ emission filling-in the line core is only detected after the 
subtraction of the low-activity template. In both the H$\alpha$ and H$\beta$ regions, the photospheric absorption lines are mostly removed by the subtraction.

Whenever a veiling $r>0$ was found, it has been introduced in the low-activity template before the subtraction, following Eq.~(\ref{eq:veiling}), so as 
to reproduce the photospheric lines of the target.
However, the $EWs$ reported in Table~\ref{Tab:gammavel} and \ref{Tab:ChaI} and stored in the GESviDR1Final catalogue are not corrected 
for veiling. To obtain the corrected values, they must be multiplied by $(1+r)$. 
 
\begin{figure*}[ht]  
\begin{center}
\includegraphics[width=8.6cm]{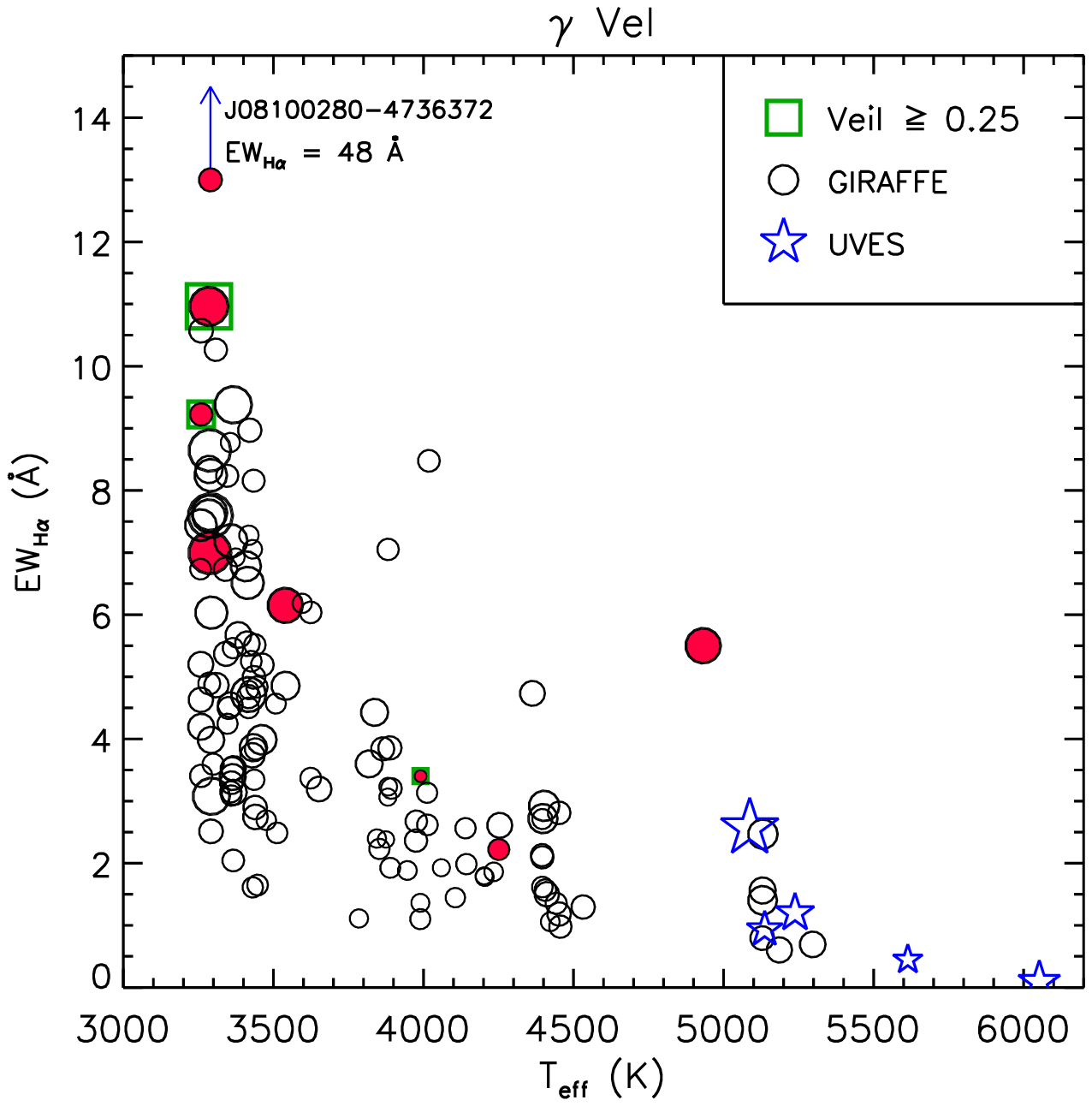}
\includegraphics[width=8.6cm]{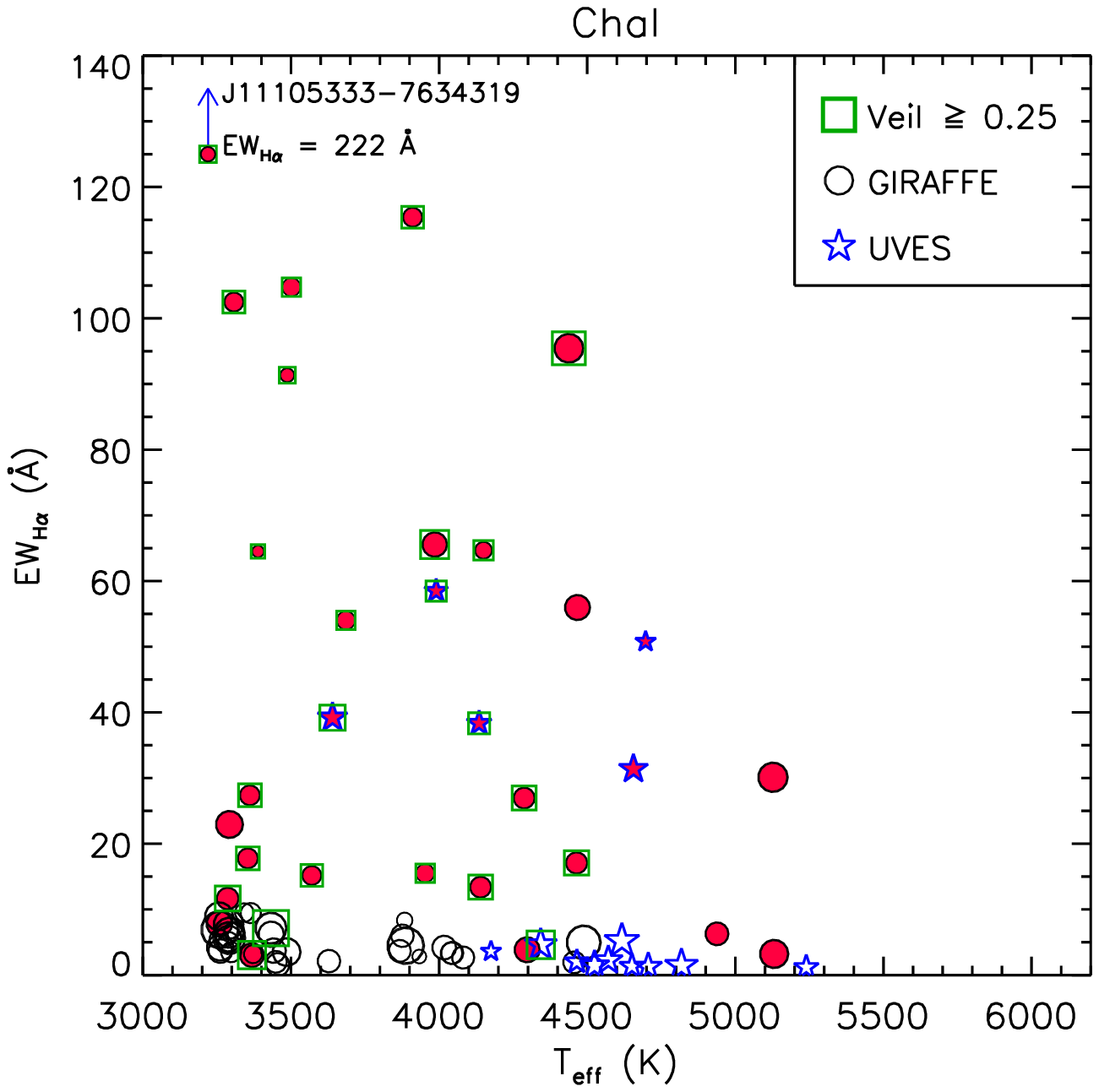}
\vspace{-.3cm}
\caption{Net H$\alpha$ equivalent width versus $T_{\rm eff}$ of the $\gamma$~Vel ({\it left panel}) and Cha~I ({\it right panel}) members 
observed with GIRAFFE and UVES. The symbol size scales with the $v \sin i$. Filled symbols denote the accretor candidates (10\%$W_{\rm H\alpha}>270$ km\,s$^{-1}$),
while the targets with a significant amount of veiling ($r \ge 0.25$) are enclosed into open squares. 
The two arrows represent the targets with $EW_{\rm H\alpha}$ out of the range. 
}
\label{fig:EW_Teff}
 \end{center}
\end{figure*}

$EW_{\rm H\alpha}$ is plotted as a function of $T_{\rm eff}$ for members of $\gamma$~Vel and Cha~I in Fig.~\ref{fig:EW_Teff}.
The largest $EW_{\rm H\alpha}$ are observed for cooler stars, due to contrast effects (i.e. the H$\alpha$ emission stands 
out against a low continuum level).  This behaviour is commonly observed in young clusters and associations \citep[see, e.g.][]{Staufferetal1997,Krausetal2014}. 

A better diagnostic of chromospheric activity is the line surface flux (indicator of radiative losses) that can be derived from the net line equivalent width as
\begin{eqnarray}
F_{\rm H\alpha} & = & F_{6563}EW_{\rm H\alpha} \\  
F_{\rm H\beta}  & = & F_{4861}EW_{\rm H\beta}\,, 
\label{eq:Fha_Fhb}
\end{eqnarray}
{\noindent where $F_{6563}$ and $F_{4861}$ are the continuum surface fluxes at the H$\alpha$ and H$\beta$ wavelengths, respectively, and
are evaluated from the NextGen synthetic low-resolution spectra \citep{hauschildtetal1999} at the stellar temperature and surface gravity provided 
by the GES consortium. }

 As a further activity index, we have also calculated the ratio of the chromospheric emission in the H$\alpha$ line to the total bolometric emission:
\begin{equation}
R'_{\rm H\alpha}  =  L_{\rm H\alpha}/L_{\rm bol} = F_{\rm H\alpha}/(\sigma T_{\rm eff}^4). 
\label{eq:Rha}
\end{equation}
 The behavior of the activity indicators as a function of stellar parameters is described in Sects.~\ref{sec:flux_temp_veil} and \ref{sec:balmer_decrement}.

\subsection{Mass accretion rate diagnostics}
\label{sec:accretion}

We considered as mass accretion rate ($\dot M_{\rm acc}$) for the members of both clusters the values reported by the GES consortium 
in the iDR1 (see \citealt{lanzafameetal2014}), which are based on the measurements of the 10\%$W_{\rm H\alpha}$ performed 
on the observed H$\alpha$ profiles, without subtracting the low-activity template. The values of $\dot M_{\rm acc}$ were computed using the \cite{nattaetal2004} relationship:
\begin{equation}
\log \dot M_{\rm acc}^{10\%W} = -12.89(\pm0.3) + 9.7(\pm0.7)10^{-3} 10\%W_{\rm H\alpha}, 
\label{eq:Natta}
\end{equation}
with 10\%$W_{\rm H\alpha}$ in km\,s$^{-1}$ and $\dot M_{\rm acc}$ in $M_{\odot}$\,yr$^{-1}$. Typical errors in $\log \dot M_{\rm acc}$ from this relation 
are about $0.4-0.5$ dex. 
\citet{nattaetal2004} provided this relation for objects with 10\%$W_{\rm H\alpha}>200$ km\,s$^{-1}$, corresponding to $\log \dot M_{\rm acc}\sim -11$. For this reason, 
the GES  data contains $\dot M_{\rm acc}$ for stars with 10\%$W_{\rm H\alpha}>$200 km\,s$^{-1}$, but only the objects that meet the most restricted 
criterion, 10\%$W_{\rm H\alpha}>270$ km\,s$^{-1}$ \citep{whitebasri2003}, are considered as accretor candidates.

An independent way of deriving the mass accretion rate is based on the total energy flux in emission lines. We used the empirical 
relations between accretion luminosity ($L_{\rm acc}$) and the luminosity in the H$\alpha$ line ($L_{\rm H\alpha}$), which were recently derived by \cite{alcalaetal2014} 
from X-Shooter@VLT data to estimate $L_{\rm acc}$. The line luminosity was calculated as $L_{\rm H\alpha} = 4 \pi R_\star^2 F_{\rm H\alpha}$, where the stellar 
radius ($R_\star$) was derived from the analysis of the SEDs (Sect.\,\ref{sec:sed}), while the surface flux ($F_{\rm H\alpha}$) was obtained 
using the net  EW of the H${\alpha}$ line, as described in Sect.\,\ref{sec:halpha_hbeta_flux}. 
Unlike several previous works, we have decided to use the net H${\alpha}$ EW, where we have removed the contribution of the
photospheric line absorption, to have a single diagnostic for both the chromospheric emission and accretion, which are simultaneously investigated. This choice 
allows us to treat properly the stars with a faint emission or only a filled-in line core. However, for the spectra showing the line as a pure emission 
feature above the continuum, we have also measured the EW of the H$\alpha$ line without subtracting the low-activity template.  
We found a negligible flux difference (within 0.1 dex), between EWs from subtracted and unsubtracted spectra, for all the accreting objects, suggesting that the
flux calculated with the net EWs can be safely used in comparison with previous works.

The mass accretion rate ($\dot M_{\rm acc}$) was then derived 
from $L_{\rm acc}$ using the relationship by \cite{hartmann1998}: 
\begin{equation}
\dot M_{\rm acc}^{EW} = \left(1 - \frac{R_\star}{R_{\rm in}}\right)^{-1} \frac{L_{\rm acc} R_\star}{G M_\star}, 
\label{eq:Hartmann}
\end{equation}
where the stellar mass $M_\star$ for each star was estimated from the theoretical evolutionary tracks, as described in Sect.~\ref{sec:HRD}, 
and the inner-disk radius $R_{\rm in}$ was assumed to be $R_{\rm in}=5R_\star$ (\citealt{hartmann1998}). Contributions to the error budget on 
$\dot M_{\rm acc}$ include uncertainties on stellar mass, stellar radius, inner-disk radius, and $L_{\rm acc}$. Assuming 
mean errors of $\sim 0.15\,M_\odot$ in $M_\star$ and $\sim 0.1\,R_\odot$ in $R_\star$, 5--10\% as relative error in $EW_{\rm H\alpha}$, 
10\% in the continuum surface flux at the H$\alpha$ line used for deriving $F_{\rm H\alpha}$, and the uncertainties in the relationships by \cite{alcalaetal2014}, 
we estimate a typical error in $\log \dot M_{\rm acc}$ of $\sim 0.5$ dex. 

The use of the H$\alpha$ EW allows us to define as ``confirmed accretors'' those objects which fulfill the requirements proposed by
\citet[][see their Fig.~7]{whitebasri2003} and based on both H$\alpha$ EW and $SpT$. Adopting these criteria,  we identified
26 and 3 accretors in Cha~I and $\gamma$~Vel, respectively. Similar results are found adopting the selection 
criteria proposed by \citet[][see their Fig.~5]{Barrado2003}. 
We remark that 24 out of 26 accretors in Cha~I were classified as flat or Class\,II IR sources by  \cite{manojetal2001} and \cite{luhmanetal2008}. For the two remaining objects 
no IR classification is available in the literature. We classified 5 Cha~I and 3 $\gamma$~Vel members, which are close to the border line proposed by \citet{whitebasri2003} or \citet{Barrado2003}, as ``possible accretors''. 
All but one of the five possible accretors in Cha~I are Class\,II objects, while the source J11071915$-$7603048 is a Class\,III star \citep{luhmanetal2008} with 
10\%$W_{\rm H\alpha}\sim370$ km\,s$^{-1}$, $v \sin i \sim 10$ km\,s$^{-1}$ and $EW_{\rm H\alpha}=15.2$\,\AA ~(see Table~\ref{Tab:ChaI}). 
Two out of our eight  confirmed/possible accretors in $\gamma$~Vel are reported as Class\,II objects by \citet{Hernandez2008}, while all the remaining six 
sources are all classified as Class\,III.

\begin{figure}[hb] 
\begin{center}
\includegraphics[width=9cm]{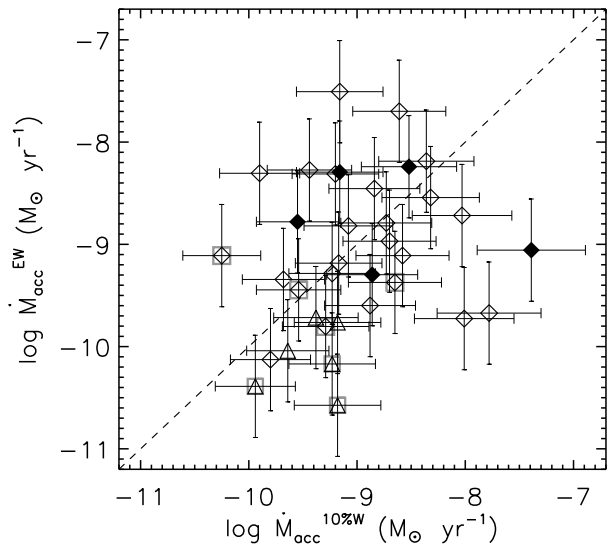}
\vspace{-.3cm}
\caption{Accretion rates from $EW_{\rm H\alpha}$ versus accretion rates from $10\%W_{\rm H\alpha}$ for Cha~I (diamonds) and $\gamma$~Vel 
(triangles) stars. Empty and filled symbols refer to GIRAFFE and UVES data, respectively. Squares show the possible accretors.
The dashed line is the one-to-one relation.
}
\label{fig:Macc_comparison}
 \end{center}
\end{figure}

In Fig.~\ref{fig:Macc_comparison}, the comparison of the two accretion rate estimates is shown. 
 The difference between the two determinations of $\dot M_{\rm acc}$  for a given object is quite large ($\sim 0.8$ dex for Cha~I and $\sim 0.7$ dex for 
$\gamma$~Vel, on  average). Similar results were also found by \cite{costiganetal2012} who studied the variability of mass accretion in a sample of 10 stars in Cha I. 
The same authors suggest that the 10\%$W_{\rm H\alpha}$ does not give reliable estimates of average accretion rates, especially when single-epoch observations 
were performed. 
The  inconsistencies we found between the two $\dot M_{\rm acc}$ determinations may also be due to the effects of H$\alpha$  extra-absorption by stellar winds 
on the  emission line profile produced by the accretion flow, and to line emission not due to accretion, which can affect the 10\% width and the H${\alpha}$ EW in 
a very different way. For instance, an extra absorption wing that produces a strongly asymmetric or a P-Cygni profile could cause an underestimate
of the 10\% width  much larger than for the H$\alpha$ EW. A Spearman's rank correlation analysis \citep{Pressetal1992} applied to the Cha\,I data 
provides a coefficient $\rho=0.10$ with a significance of its deviation from zero $\sigma=0.61$ which testifies the large data scatter.

For our Cha~I data, most of the spread is due 
to seven stars showing differences in the accretion rates larger than 1 dex. In particular, for the 
three stars with $(\log \dot M_{\rm acc}^{10\%W} - \log \dot M_{\rm acc}^{EW}) > +1.0$ dex, namely J11092379$-$7623207, J11071206$-$7632232, and 
J11075809$-$7742413 (with $v \sin i$ of some km\,s$^{-1}$), the difference is most probably due to the presence of wide wings and/or strong central reversals 
in their spectra at the epoch of our observations. This overestimates the 10\%$W_{\rm H\alpha}$ and, therefore, the mass accretion rate 
derived from this diagnostic.  
For the four stars with $(\log \dot M_{\rm acc}^{10\%W} - \log \dot M_{\rm acc}^{EW}) < -1.0$\,dex, the difference between the two $\dot M_{\rm acc}$ values could 
be due  instead to  the \cite{nattaetal2004} relation in this range of values.
In fact, as pointed out by \cite{alcalaetal2014}, for objects with 10\%$W_{\rm H\alpha} < 400$ km\,s$^{-1}$, the \cite{nattaetal2004} relation tends 
to underestimate $\dot M_{\rm acc}$ by $\sim0.6$ dex  with respect to the determinations coming from primary diagnostics such as the 
continuum-excess modelling;  however, the differences may be up to about one order of magnitude.  Indeed, the stars 
in our sample with $(\log \dot M_{\rm acc}^{10\%W} - \log \dot M_{\rm acc}^{EW}) < -1.0$ dex have $270 < 10\%W_{\rm H\alpha} \ltsim 440$ km\,s$^{-1}$. 
\cite{herczeghillenbrand2008}, \cite{fangetal2009}, and \cite{costiganetal2012} reported similar findings in Taurus, L1641, and Cha~I, respectively. 

Concerning the data that we acquired for $\gamma$~Vel, only the possible accretor J08105600$-$4740069 shows 
$(\log \dot M_{\rm acc}^{10\%W} - \log \dot M_{\rm acc}^{EW}) > 1 $ dex. 		 
This source displays wide H$\alpha$ wings, $10\%W_{\rm H\alpha}$ close to 400 km\,s$^{-1}$, and a moderate rotation rate
($v \sin i \sim 15$ km\,s$^{-1}$). It is also a Class\,II object according to \citet{Hernandez2008}.

\section{Results and Discussion}
\label{sec:discussion}

\subsection{Projected rotation velocity}
\label{sec:vsini}

The availability of a large dataset of cluster members with measured projected rotational velocity allows us to investigate the 
 distribution of stellar rotation rates and their dependence on fundamental stellar parameters.

\begin{figure*}  
\begin{center}
\includegraphics[width=8.8cm]{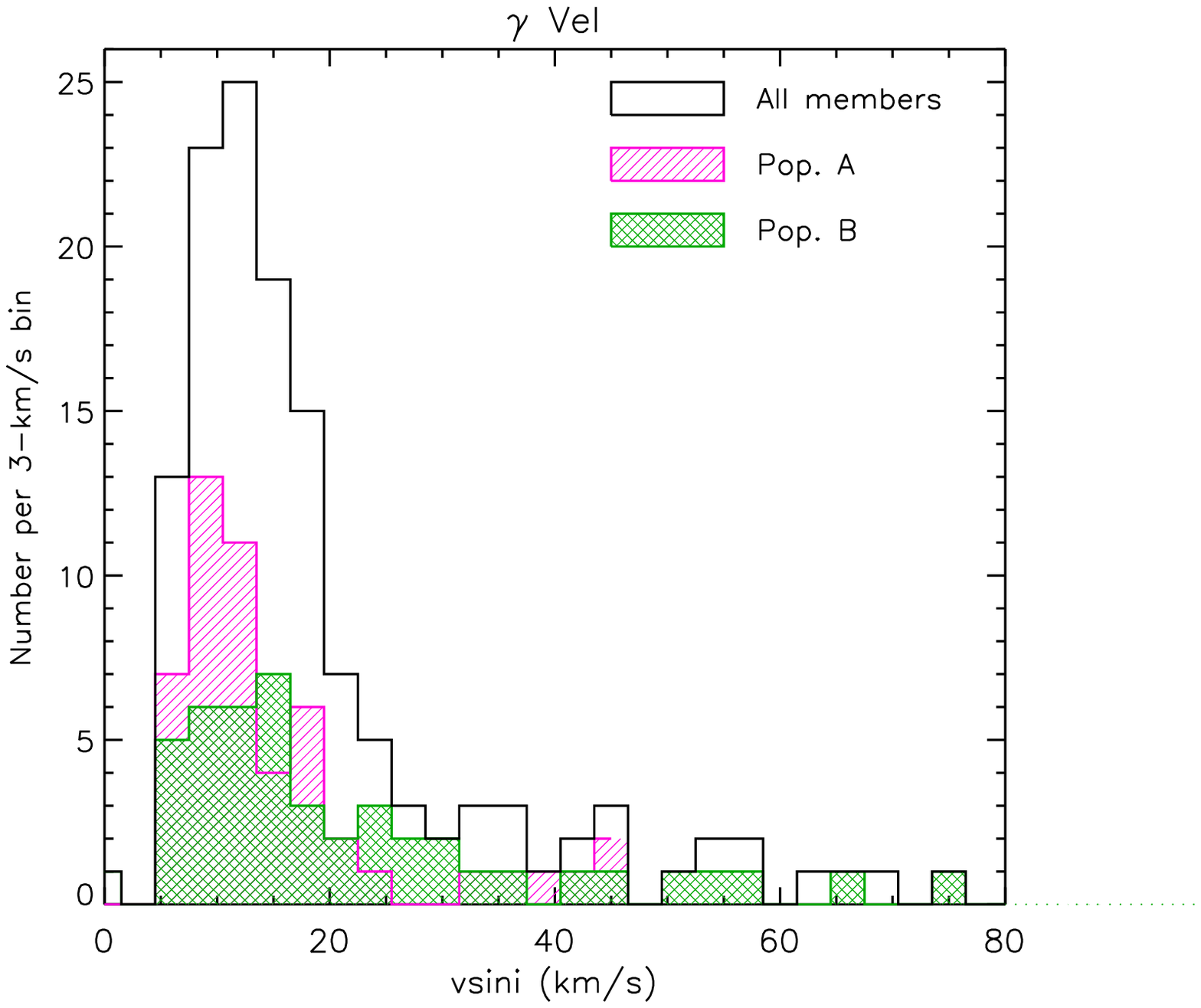}
\includegraphics[width=8.8cm]{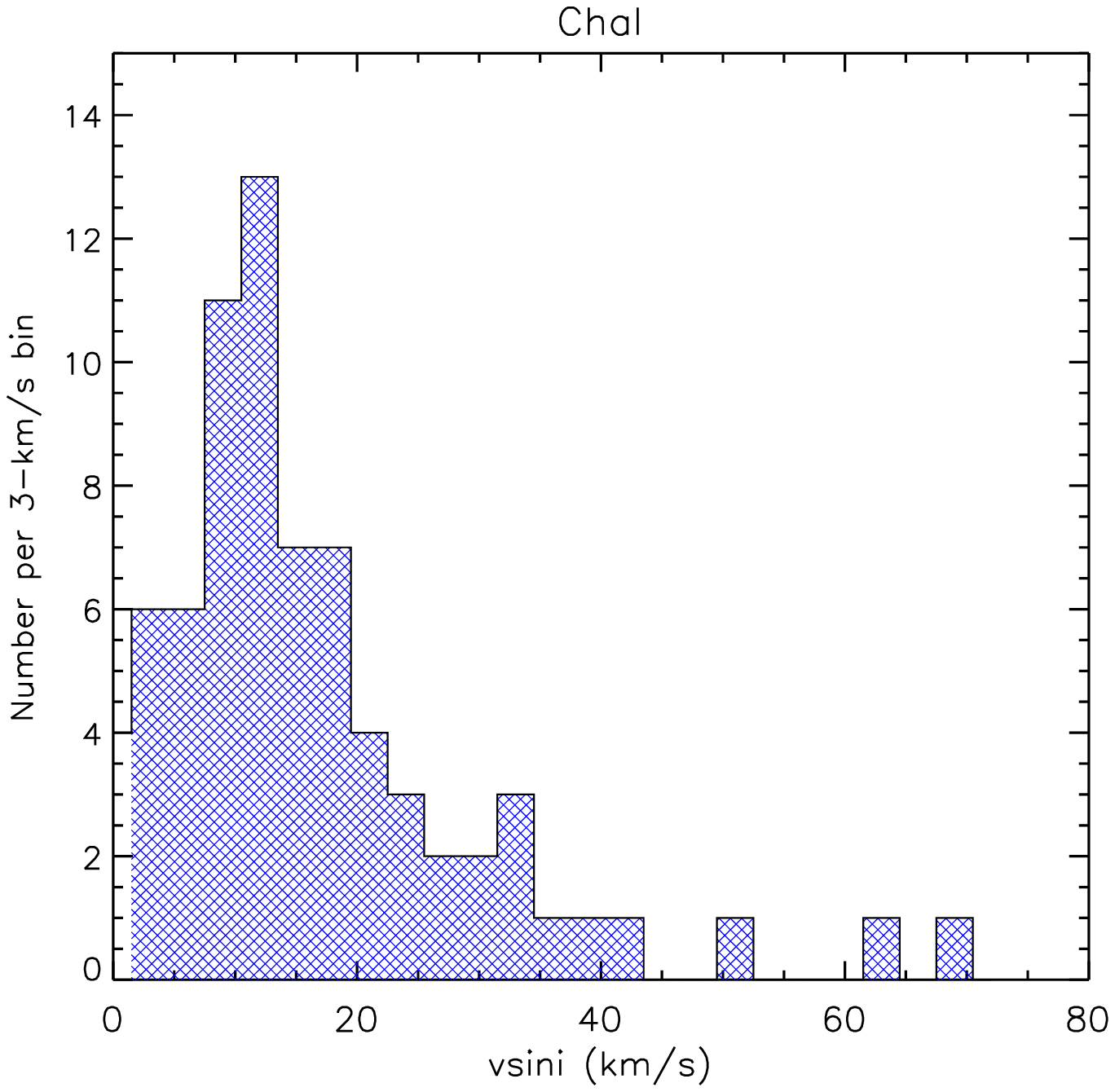}
\vspace{-.5cm}
\caption{{\it Left panel:} The distribution of $v \sin i$ for the members of $\gamma$~Vel (empty histogram) showing a main peak 
centered at about 10 km\,s$^{-1}$ with a tail towards faster rotators. The two kinematic subsamples A and B identified by \citet{jeffriesetal2014}
display slightly different distributions (hatched and filled histograms), with higher frequency of faster rotators for the population B. 
{\it Right panel:} The distribution of $v \sin i$ for the members of Cha~I, peaked at about 10 km\,s$^{-1}$.}	
\label{fig:vsini_distr}
 \end{center}
\end{figure*}

Figure~\ref{fig:vsini_distr} shows the distribution of $v\sin i$ for both clusters. 
The $v\sin i$ was measured for all 132 GIRAFFE members of the $\gamma$~Vel cluster according to the three criteria adopted by \citet{jeffriesetal2014}, 
namely CMD, lithium line, and $RV$. We also have $v\sin i$ determinations for the eight late-type members of $\gamma$~Vel observed with UVES. 
Two of them have a $RV$ not compatible with the cluster, but they fulfill all the other criteria and are considered as members by \citet{spinaetal2014a}. 
Two of these eight UVES targets have been also observed with GIRAFFE in different observing blocks, but in this study we considered for them the UVES data. 
The $v\sin i$ distribution (left panel in Fig.~\ref{fig:vsini_distr}) displays a main peak at about 10 km\,s$^{-1}$ with a tail toward faster rotators. 
Despite the blurring of the distribution produced by the inclination angles, compared to a rotation period distribution, its appearance 
is consistent with a mixture of stars that have spun up and others that have maintained a slow rotation rate likely due to efficient disk locking.  
We  constructed the $v\sin i$ distributions for the stars that can be unambiguously associated with each of the two kinematical populations identified by 
\citet{jeffriesetal2014}  and clearly  revealed by the double-peaked distribution of the radial velocities (see Fig.~\ref{fig:vrad_distr_gammaCha_G}). 
The population A, centered at about 16.7 km\,s$^{-1}$ with an intrinsic dispersion $\sigma_{\rm A}=0.34$\,km\,s$^{-1}$ is found to be  older by about 1--2\,Myr than 
the component B, which is centered at 18.8 km\,s$^{-1}$ and shows a wider dispersion ($\sigma_{\rm B}=1.60$\,km\,s$^{-1}$).
As already noted by the same authors, the stars in the population B  tend to rotate 
faster, on average, than those of the population A.  A two-sided Kolmogorov-Smirnov (KS; \citealt{Pressetal1992}) test of the cumulative $v\sin i$ distributions of the two
populations reveals a significant difference, the significance level resulting to be $P_{\rm KS}=0.03$.
 This behaviour cannot be attributed to the age difference between the two groups, which is too
small in comparison with the typical times of rotation evolution and is more likely related to different environmental conditions during their early
life.  
The massive binary system $\gamma^2$\,Vel seems to be slightly younger than the low-mass stars of the population A  \citep{jeffriesetal2014}.
Thus, these stars may not have been affected by the strong radiation field and stellar wind from $\gamma^2$\,Vel during the first few Myr of their life,
while  the population B might have experienced such an effect. As a result, the disks around the members of population B could have been dispersed earlier than those of the 
population A, with a shorter disk-locking effect and a faster spin up.  

For Cha~I,  the distribution displays a peak around 10 km\,s$^{-1}$,  which is narrower than that of $\gamma$\,Vel,  and a non negligible fraction of relatively 
fast rotators (up to $\sim$\,40 km\,s$^{-1}$).  A KS test of the $\gamma$\,Vel and Cha\,I $v\sin i$ distributions shows only a marginal difference
 ($P_{\rm KS}=0.36$). This is in agreement with the results of studies of the evoultion of stellar rotation \citep[e.g.,][]{messinaetal2010,spadaetal2011} that show only a 
moderate increase of the average rotation rate between the ages of these two clusters.

\subsection{H$\alpha$ flux}
\label{sec:flux_temp_veil}

\begin{figure*}  
\begin{center}
\includegraphics[width=8.8cm]{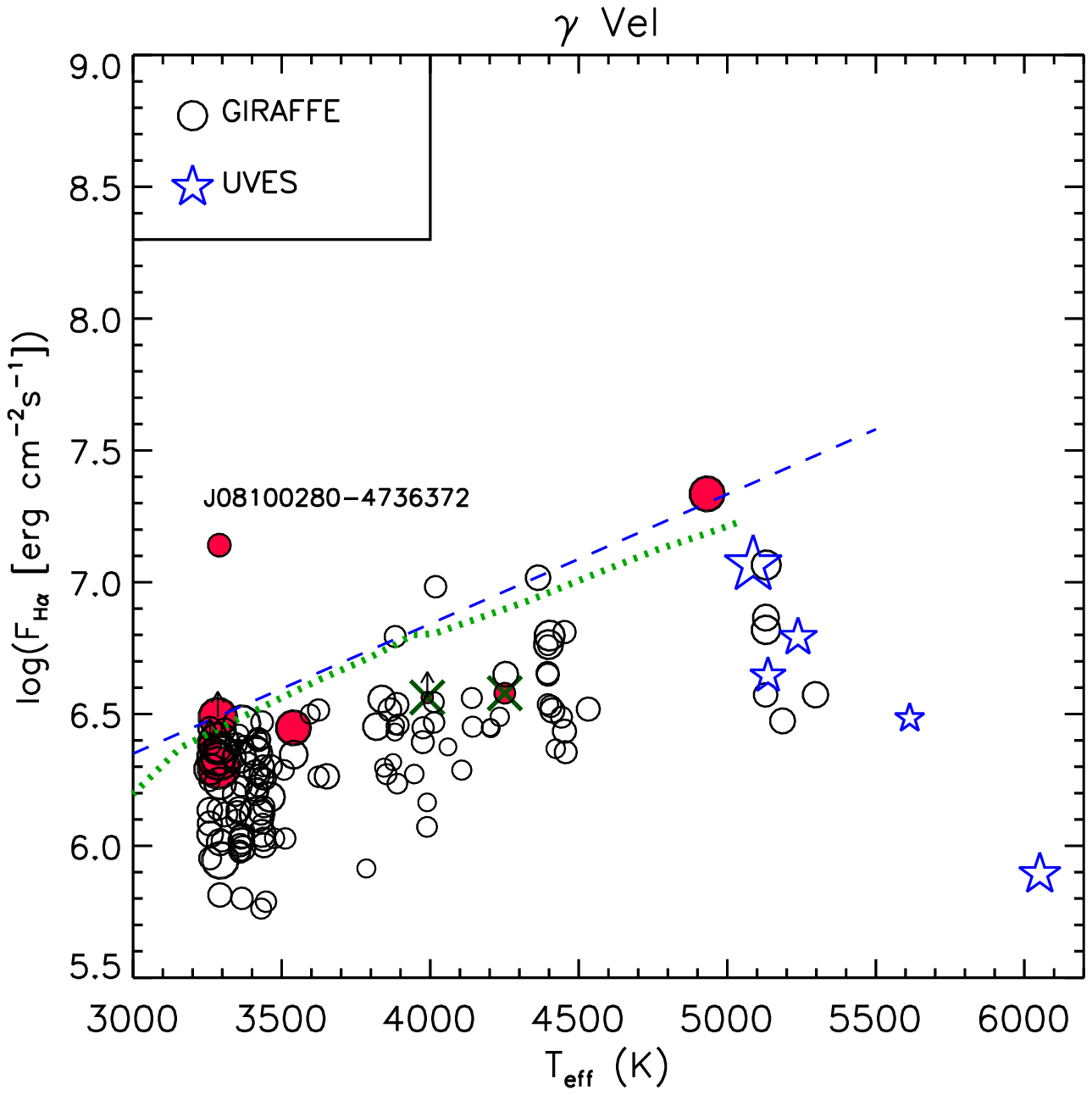}
\includegraphics[width=8.8cm]{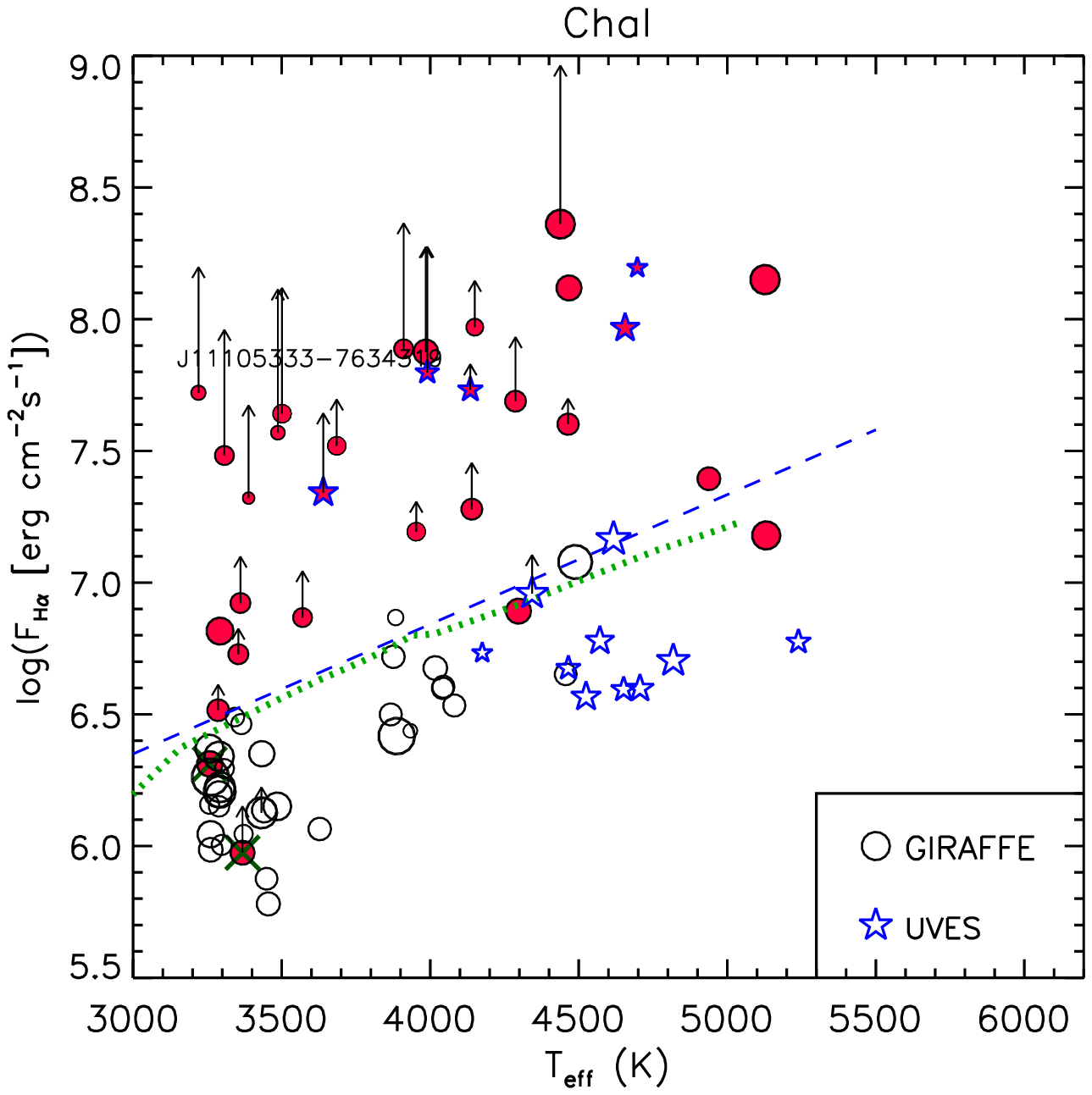}
\vspace{-.3cm}
\caption{H$\alpha$ flux versus $T_{\rm eff}$ for the $\gamma$~Vel ({\it left panel}) and the Cha~I ({\it right panel}) members observed with 
GIRAFFE and UVES. The symbol size scales with the $v \sin i$. The accretor candidates (10\%$W_{\rm H\alpha}>270$ km\,s$^{-1}$)  
are denoted with filled symbols and typically have a flux larger than the other stars.
The candidates that have been rejected on the basis of the $EW_{\rm H\alpha}$ criterion (Sect.~\ref{sec:accretion}) are marked with crosses. 
The flux values corrected for veiling by the factor $(1+r)$ are denoted by arrowheads.
In each box, the dashed straight line is drawn to follow the upper envelope of the sources without accretion, while the dotted line is
the saturation criterion adopted by \citet{Barrado2003} to separate classical from weak T Tauri.
}
\label{fig:flux_Teff_G_gamma2velChaI}
 \end{center}
\end{figure*}

\begin{figure*}  
\begin{center}
\includegraphics[width=8.8cm]{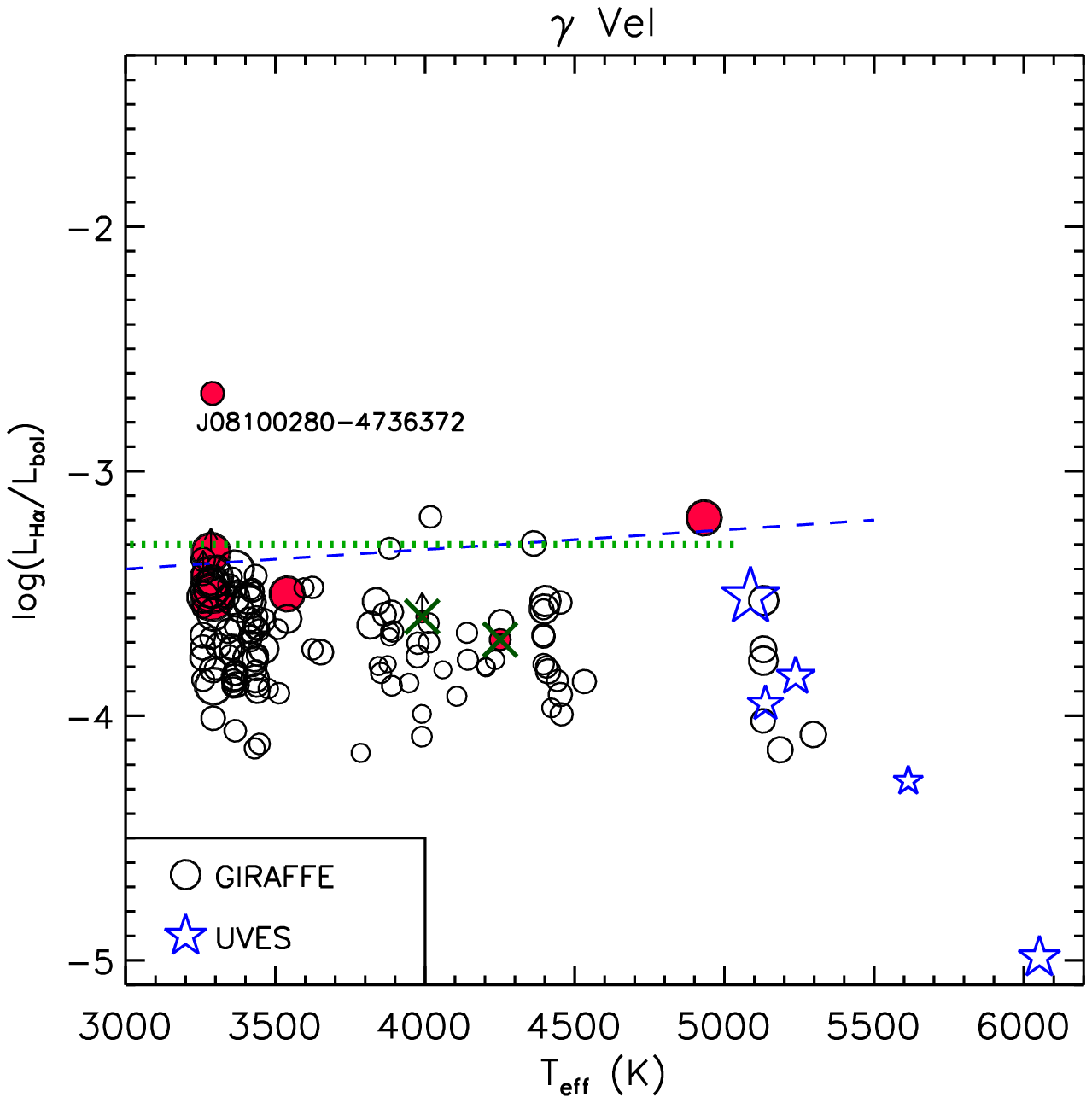}
\includegraphics[width=8.8cm]{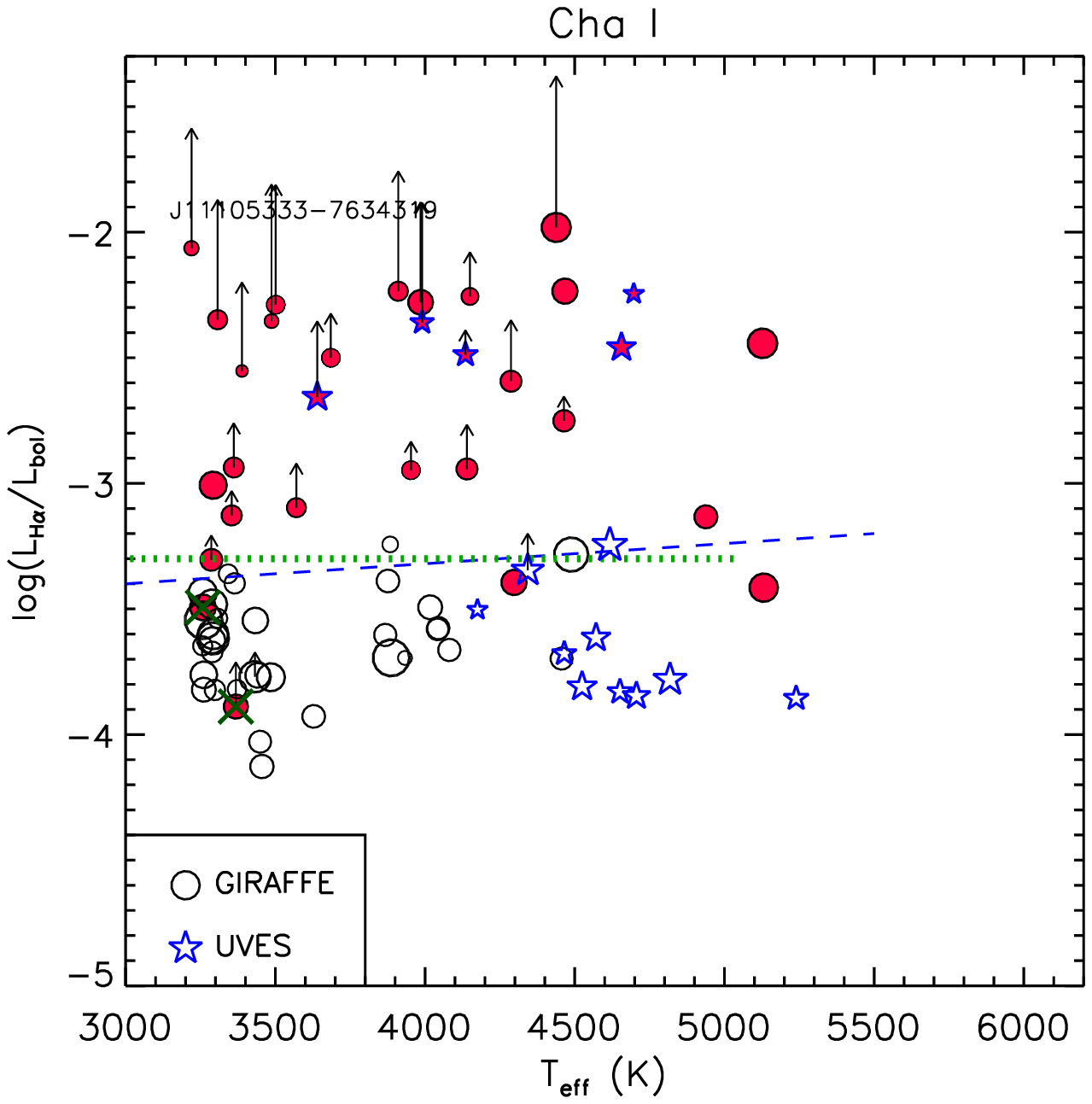}
\vspace{-.3cm}
\caption{ $R'_{\rm H\alpha}$ versus $T_{\rm eff}$ for the $\gamma$~Vel ({\it left panel}) and the Cha~I ({\it right panel}) members observed with 
GIRAFFE and UVES. The meaning of the symbols, arrowheads, and dashed/dotted lines is as in Fig.~\ref{fig:flux_Teff_G_gamma2velChaI}. 
}
\label{fig:Lha_Lbol}
\end{center}
\end{figure*}

In Fig.~\ref{fig:flux_Teff_G_gamma2velChaI} we show the H$\alpha$ surface flux as a function of the effective temperature. 
This figure clearly shows that the nearly exponential behavior displayed by $EW_{\rm H\alpha}$ as a function of $T_{\rm eff}$ (Fig.~\ref{fig:EW_Teff})  disappears when the flux is used.
 In this figure we do not use squares to enclose the stars with a veiling $r\geq 0.25$, as we did in Fig.~\ref{fig:EW_Teff}, but we 
display with arrows the flux values obtained by correcting the $EW$s for the dilution caused by the veiling, i.e. multiplying by the factor $(1+r)$.

We have drawn, in both panels of Fig.~\ref{fig:flux_Teff_G_gamma2velChaI} a dashed straight line that demarcates the
 domain of accretors from that of chromospherically active stars. This ``dividing line'', which is empirically defined by the upper boundary of the  
chromospheric fluxes (empty symbols) of stars in both clusters,  is expressed by:
\begin{equation}
\label{eq:dividing}
\log F_{\rm H\alpha}=6.35 + 0.00049(T_{\rm eff}-3000)\,. 
\end{equation}

For comparison, we have also overplotted in Fig.~\ref{fig:flux_Teff_G_gamma2velChaI}  the ``saturation limit'' adopted by \citet{Barrado2003} 
to separate classical from weak T~Tauri stars, where we adopted the $SpT$--$T_{\rm eff}$ calibration of \citet{PecautMamajek2013}. The two boundaries 
are very close, especially for the coolest stars, where the subtraction of the non-active template has a negligible effect on the $F_{\rm H\alpha}$ due to both the 
faint photospheric absorption (compared to usually strong line emissions) and the low continuum flux.

The behaviour of $R'_{\rm H\alpha}$ versus $T_{\rm eff}$  for both clusters is displayed in Fig.~\ref{fig:Lha_Lbol}, where a much flatter
trend appears. The accretor candidates lie in the upper part of these plots too.  The dividing line for this activity index, overplotted with a 
dashed line, is given by:
\begin{equation}
\label{eq:dividing_Lbol}
\log R'_{\rm H\alpha}=-3.4 + 0.00008(T_{\rm eff}-3000)\,. 
\end{equation}
  
We note that the average value of this line is close to the saturation limit, $\log R'_{\rm H\alpha}=-3.3$, adopted by \citet{Barrado2003} as the
boundary between accreting and non-accreting objects, which is also displayed in Fig.~\ref{fig:Lha_Lbol}.

In the case of $\gamma$~Vel, most stars have $F_{\rm H\alpha}$ close to the maximum values found by \citet[][see their Fig.\,7]{martinez-arnaizetal2011} 
for stars with X-ray luminosity in the saturated regime.  Moreover, the fluxes do not seem to correlate with the $v\sin i$, as indicated by the Spearman's 
rank correlation coefficient $\rho=0.057$ and by the two-sided significance of its deviation from zero $\sigma=0.519$, even rejecting the few accretors. 
A higher coefficient ($\rho=0.452$) with a $\sigma=7.6\times 10^{-8}$ is found instead for $\log R'_{\rm H\alpha}$ versus $v\sin i$. This suggests that most stars 
have already reached the saturation of magnetic activity, while the remaining objects are likely contributing to this residual correlation
which is best detected in the $\log R'_{\rm H\alpha}$ diagnostic. Similar results are found for Cha\,I when the accreting objects are disregarded.

 Eight out of the 140 UVES+GIRAFFE members of the $\gamma$\,Vel cluster are accretor candidates (10\%$W_{\rm H\alpha}>270$\,km\,s$^{-1}$), 
but we only confirmed three accretors. The percentage of accretors is then 2\,\% or, at most, 4\,\% if we consider also the possible accretors.  
All these objects fall above or close to the dividing line, while two candidates, namely  J08103074$-$4726219 and J08104649$-$4742216, lie well below 
this boundary (see Figs.\,\ref{fig:flux_Teff_G_gamma2velChaI} and \ref{fig:Lha_Lbol}, left panels).
The first one has 10\%$W_{\rm H\alpha}\sim280$ km\,s$^{-1}$ and was finally classified as a non-accretor. 
The second one shows a large 10\%$W_{\rm H\alpha}$ uncertainty 
($\sim 30 \%$), due to bad quality of the spectrum and cannot be considered as an accretor according to the criteria adopted in Sect.~\ref{sec:accretion}.
 Both are reported as Class\,III sources by \citet{Hernandez2008} based on their SED.

All the members of Cha~I  with 10\%$W_{\rm H\alpha}>270$\,km\,s$^{-1}$ fall above or very close to the dividing line with only
two exceptions. One of these two stars is J11085242$-$7519027 ($T_{\rm eff} \sim 3400$ K, 10\%$W_{\rm H\alpha}\sim 272$\,km\,s$^{-1}$), which
we finally do not classify as an accretor, while the other, namely J11122441$-$7637064 ($T_{\rm eff} \sim 5100$ K and 10\%$W_{\rm H\alpha}\sim 380$\,km\,s$^{-1}$), 	
is defined as an accretor. Moreover, J11122441$-$7637064 was previously classified as a classical T Tauri on the basis of {\it Spitzer} photometry 
\citep[e.g.,][]{Wahhaj2010} and \citet{Nguyen2012} reported a value of  10\%$W_{\rm H\alpha}$ (381\,km\,s$^{-1}$) very close to our measurement. 
\citet{luhman2007} found a lower effective temperature for it 
($T_{\rm eff} \sim 4660$\,K) from low-dispersion spectra. If we adopt this temperature, we obtain a flux $\log F_{\rm H\alpha}\simeq 7.0$ (in cgs units) 
that leads the star slightly closer to the dividing line.  However, this star is a visual binary with a companion at about 
2$\arcsec$ \citep{daemgenetal2013}, whose light could have contaminated the GIRAFFE spectrum. 

As mentioned in Sect.~\ref{sec:spec_analysis}, the veiling 
was taken as a free parameter only for the stars with a strong and broad H$\alpha$ emission, which was considered as  the main requirement for the
pre-selection of accretor candidates within the GES. We found no star in $\gamma$~Vel with $r>0.25$. Among the eight stars 
with a veiling detection, one is an accretor and other two are possible accretors according to our definition in Sect.~\ref{sec:accretion}.

\begin{figure}  
\begin{center}
\includegraphics[width=8.8cm]{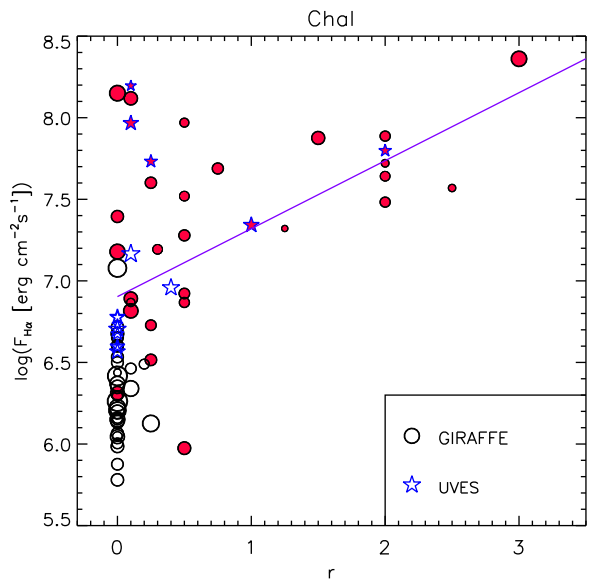}
\vspace{-.3cm}
\caption{H$\alpha$ flux versus veiling for the Cha~I members observed with GIRAFFE and UVES. 
 The accretor candidates (10\%$W_{\rm H\alpha}>270$ km\,s$^{-1}$) are denoted by filled symbols, as in Fig.~\ref{fig:flux_Teff_G_gamma2velChaI}.
All the stars with a significant veiling ($r \ge 0.5$) turn out to be confirmed or possible accretors. The full line is a linear best fit to the data with $r \ge 0.25$.
}
\label{fig:flux_Teff_veil_G_ChaI}
 \end{center}
\end{figure}

In Cha~I the picture is much different. Among the 74 UVES+GIRAFFE members, 31 sources (about 42\,\%) display 10\%$W_{\rm H\alpha}>270$ km\,s$^{-1}$ 
and all are confirmed (26 sources, i.e. 35\,\%) or possible (5 sources) accretors.  Moreover, most of them lie above the line of non-accreting stars in Fig.\,\ref{fig:flux_Teff_G_gamma2velChaI} 
by about 0.5--1.0 dex. In addition,  as displayed in Fig.~\ref{fig:flux_Teff_veil_G_ChaI}, all the stars with significant veiling ($r\ge 0.5$) are accretors 
(see also Sect.~\ref{sec:macc_mass}).
Figure~\ref{fig:flux_Teff_veil_G_ChaI} also shows a positive correlation between H$\alpha$ flux and $r$,  at least for
the objects with $r \ge 0.25$ for which the Spearman's rank analysis yields a coefficient $\rho=0.58$ with a significance of $\sigma=0.003$. 
Presently, a more accurate analysis of the stellar properties versus veiling cannot be done because of the uncertainties of the veiling values for GIRAFFE
spectra due to their limited spectral range and the absence of strong photospheric lines in the HR15N setup. 

Concluding, these two clusters are different both in the accretion properties and in the emitted average H$\alpha$ line flux, with $\gamma$\,Vel 
showing less accretion signatures than Cha~I, as expected due to its older age. In particular, the line flux emitted 
by the confirmed or possible accreting objects in $\gamma$\,Vel is comparable to, or just larger than, the highest chromospheric fluxes 
emitted by the other $\gamma$\,Vel members, suggesting that most of these stars are near the end of the accretion phase.

\subsection{Balmer decrement}
\label{sec:balmer_decrement}

\begin{figure}  
\begin{center}
\includegraphics[width=8.6cm]{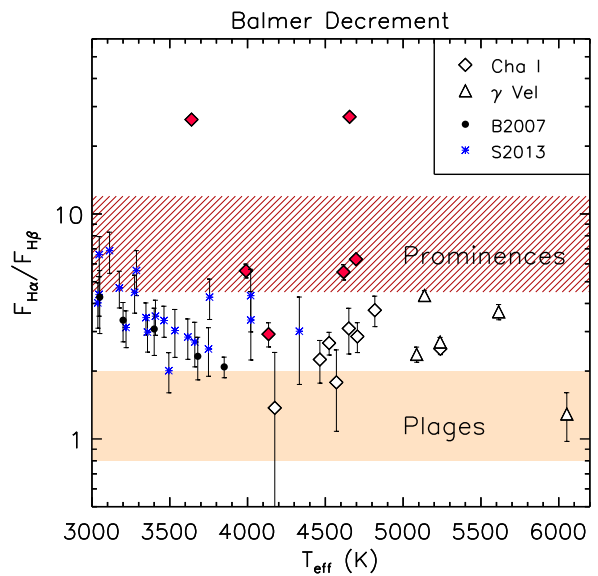}
\vspace{-2.cm}
\caption{Balmer decrement ($F_{\rm H\alpha}/F_{\rm H\beta}$) versus effective temperature for the $\gamma$~Vel (triangles) and Cha~I (diamonds) stars 
with residual emission detected both in the H$\alpha$ and H$\beta$ lines. The accretors are displayed with filled symbols.  The decrements measured for late-K and
M-type stars by \citet[][B2007]{bochanskietal2007} and \citet[][S2013]{stelzeretal2013} are overplotted with different symbols.
The range of values typical for solar
plages and for prominences are also shown by the shaded and hatched areas, respectively.
}
\label{fig:balm_decr}
 \end{center}
\end{figure}

The H$\alpha$ and H$\beta$ fluxes measured in the UVES spectra allowed us to calculate the Balmer decrement  ($F_{\rm H\alpha}/F_{\rm H\beta}$) 
that is a sensitive indicator of the physical conditions,  mainly density and temperature, in the emitting regions. 
 The Balmer decrement for the $\gamma$~Vel and Cha\,I members is plotted versus $T_{\rm eff}$ in Fig.\,\ref{fig:balm_decr}.

It is well know that the Balmer decrement for the Sun is quite low ($\sim$\,1--2)  in the optically thick plasma of plages or pre-flare active regions,
while it is much higher ($\sim$\,4.5--12) in the prominences \citep[see, e.g.][]{tandberg1967,landman1979,chester1991}.

For the very active giants or subgiants in RS~CVn binaries, a Balmer decrement in the range 3--10, i.e. significantly larger than that of solar 
active regions,  has been observed. This has been interpreted as the result of different conditions in the active regions or as the combined effect 
of plage-like and prominence-like structures  \citep[][]{hallramsey1992,chesteretal1994}.
For active main-sequence stars, a lower Balmer decrement (in the range 2.2--3.2), but still slightly larger than in solar plages, has been observed 
\citep[see, e.g.][]{frascaetal2010,frascaetal2011}. In the case of late-K and M-type stars, a Balmer decrement intermediate between solar plages 
and prominences (2--5), with an increasing trend with the decrease of $T_{\rm eff}$, has been observed  both in field dMe stars
\citep{bochanskietal2007} and in PMS Class\,III stars in regions with age of 1--10 Myr \citep{stelzeretal2013}. These data are also displayed
in Fig.\,\ref{fig:balm_decr} for comparison.

For the few chromospherically active stars members of $\gamma$~Vel and for the non-accreting stars in Cha~I we also found a Balmer decrement 
between about 2 and 5, i.e. intermediate between solar plages and prominences.  
This suggests either that the chromospheric active regions of these young stars have a different structure, mainly as regards their 
optical thickness, compared to the solar plages or that the emitted  chromospheric flux is the results of contributions from plage-like and prominence-like 
regions, the latter having a much higher Balmer decrement. 
Moreover, these data do not show any clear dependence of the Balmer 
decrement on $T_{\rm eff}$ for the G--K-type stars in 1--10\,Myr age range, unlike what is seen for M-type stars \citep{stelzeretal2013}.

The star J11064510$-$7727023 (=\object{UX~Cha}) was disregarded in this analysis because 
of unreliable $EW_{\rm H\beta}$ value due to extremely low S/N of the spectrum in the H$\beta$ region. The six accreting stars in Cha~I observed 
with UVES display instead higher Balmer decrements, from about 3 to 30, as expected from an optically thin accreting matter.

\subsection{Mass accretion rate}
\label{sec:macc_mass}

\begin{figure*}  
\begin{center}
\includegraphics[width=17cm]{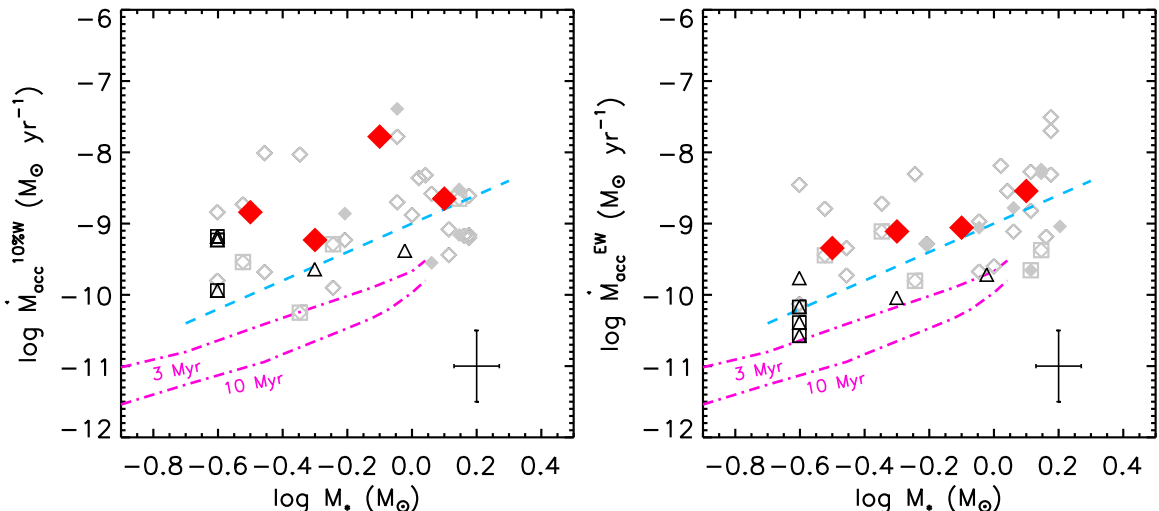}
\vspace{.1cm}
\caption{Mass accretion rate  from $10\%W_{\rm H\alpha}$ ({\it left panel}) and $EW_{\rm H\alpha}$ ({\it right panel}) versus 
stellar mass. Diamonds and triangles represent Cha~I and $\gamma$~Vel stars, where filled and empty symbols refer to UVES and GIRAFFE 
data, respectively. Squares mark the position of the possible accretors. Big red diamonds represent the median values of $\dot M_{\rm acc}$ 
for stellar masses of Cha~I members binned at $0.2 M_\odot$, where both confirmed and possible accretors were considered. The dashed line 
represents the $\dot M_{\rm acc}\propto M_\star^2$ relation, while the ``noise boundaries'' at 3~Myr and 10~Myr due to chromospheric activity 
are overplotted by dash-dotted lines (\citealt{manaraetal2013}). Mean error bars are overplotted on the right corners of both panels. 
}
\label{fig:Macc_Mass}
 \end{center}
\end{figure*}

In Fig.~\ref{fig:Macc_Mass}, the mass accretion rates measured by means of the $10\%W_{\rm H\alpha}$ and $EW_{\rm H\alpha}$ diagnostics are plotted 
as a function of the stellar mass derived from the HR diagram (see Sect.~\ref{sec:HRD}). From this figure, it is evident how all confirmed and
possible accretors in both  clusters fall above the boundaries between chromospheric emission and accretion as defined by \citet{manaraetal2013} for the ages of  
Cha~I and $\gamma$~Vel ($\sim 3$ Myr and $\sim 10$ Myr, respectively).
This means that the selection of accreting objects within the GES is reliable.

Moreover, from the same figure, it is also evident the large spread 
in accretion rates for any given mass  similar to what already found by previous studies.
Short-term (e.g., \citealt{biazzoetal2012}) and long-term variability (up to $\sim$0.5\,dex according to \citealt{costiganetal2014}) may contribute to, but do 
not explain the large vertical spread of the $\dot M_{\rm acc} - M_\star$ relationship.  Different methodologies to derive $\dot M_{\rm acc}$ 
(see, e.g., \citealt{alcalaetal2014}) may also contribute to the scatter, but the large spread (up to 3~dex) observed in the 
$\dot M_{\rm acc} - M_\star$ relation is still unexplained. What seems to be reached is the general agreement in finding 
a dependence of $\dot M_{\rm acc}$ on stellar mass with a power of $\sim 2$  (e.g., \citealt{muzerolleetal2005,herczeghillenbrand2008, alcalaetal2014}).
From Fig.~\ref{fig:Macc_Mass}, despite the large spread in $\dot M_{\rm acc}$ at each stellar mass, an increasing trend of $\dot M_{\rm acc}$ with $M_\star$ 
seems to be present, and this emerges more clearly when $\dot M_{\rm acc}$ is derived  from $EW_{\rm H\alpha}$. 
The power-law relation  $\dot M_{\rm acc}\propto M_\star^{2}$, which is also
depicted in Fig.~\ref{fig:Macc_Mass}, is not inconsistent with our data, although the scatter does not allow us to say more. A Spearman's rank 
correlation analysis yields for Cha\,I a coefficient $\rho=0.44$ with a significance $\sigma=0.01$ for $\dot M_{\rm acc}$ derived from $EW_{\rm H\alpha}$, indicating a significant positive correlation, 
while $\dot M_{\rm acc}$ is less correlated with the stellar mass ($\rho=0.26$, $\sigma=0.16$)  when it is obtained from $10\%W_{\rm H\alpha}$.

Summarizing, the mass accretion rate for the few accretors in the $\gamma$~Vel sample ranges from $\sim 10^{-11}$ to $10^{-9} M_\odot$\,yr$^{-1}$, 
while for the Cha~I members it is in the range $10^{-10} - 5 \times 10^{-8}$ $M_\odot$\,yr$^{-1}$, with a mean value of 
$\sim 2 \times10^{-9} M_\odot$\,yr$^{-1}$ at $\sim 1 M_\odot$. Considering the different ages of $\gamma$~Vel and Cha~I, 
these values are consistent with the temporal decay of mass accretion rates due to the mechanisms driving the evolution and dispersal 
of circumstellar disks \citep[see, e.g.,][]{hartmannetal1998}.
Moreover, as mentioned in Sect.~\ref{sec:flux_temp_veil}, we find a fraction of accretors of $\sim 35-42\%$ in Cha~I and 
$\sim 2-4\%$ in $\gamma$~Vel, which are in very good agreement with the results of \citet{Fedeleetal2010} based on low resolution spectra and 
consistent with the disk fractions for stellar clusters with ages similar to that of $\gamma$~Vel and Cha~I  (\citealt{ribasetal2014}).

A comparison between our mass accretion rates with those derived in the literature is presented in Appendix\,\ref{sec:macc_literature}.

\section{Summary and Conclusions}
\label{sec:conclusions}

In this paper we used the dataset provided by the GES consortium to study the chromospheric activity and accretion properties of the $\gamma$\,Vel and
Cha\,I regions. Our findings can be summarized as follows:

\begin{itemize}
\item GIRAFFE spectra acquired within the GES survey with a $S/N > 20$ for stars in the young associations $\gamma$~Vel and 
Cha~I form statistically significant samples for the analysis of $v \sin i$.
\item The $v \sin i$ distribution for the members of  $\gamma$~Vel appears asymmetric with a main peak at about 10 km\,s$^{-1}$ and a broad tail extending 
towards fast rotators. This suggests the presence of both stars that are at the end of the disk-locking phase and are still rotating rather slowly and others that 
have started to spin-up while contracting and approaching the ZAMS.   Some indication of a distinction between the A and B kinematical subsamples 
discovered by \citet{jeffriesetal2014} emerges from the $v\sin i$ data.
\item We found no clear dependence of the chromospheric H$\alpha$ flux on $v \sin i$. Only a hint of correlation with the $v\sin i$
emerges instead for $\log R'_{\rm H\alpha}$, i.e. the line flux normalized to the bolometric one. This very weak dependence on $v\sin i$
may be due to activity saturation for most of the non-accreting stars, as witnessed by the high chromospheric fluxes that are comparable to those typical for stars 
in the saturated regime.	
\item A low fraction ($\sim 2-4\%$) of $\gamma$~Vel members display mass accretion, while a much larger percentage 
($\sim 35-42\%$) was found for Cha\,I. 
This is an expected result on the basis of the quick dissipation of the disks after their typical lifetime of 5--7 Myr  
and suggests that $\gamma$~Vel is right at the end of the accretion phase.  
\item Accreting and active stars occupy two different regions in a $T_{\rm eff}$--$F_{\rm H\alpha}$ diagram and we propose a simple  criterion for distinguishing 
them, which is, however, well consistent with previous findings \citep[e.g., ][]{Barrado2003}. 
The few stars around the dividing line in our plots are possibly near the end of their accretion phase or have very high chromospheric fluxes. 
\item The Balmer decrement ($F_{\rm H\alpha}/F_{\rm H\beta}$) was calculated for the stars observed with UVES, where the setup included both
H$\alpha$ and H$\beta$. 
In the case of the active stars in $\gamma$~Vel and the non-accreting members of Cha\,I, we found values in the range 2--5, that are slightly larger than those 
observed in solar plages, as already found in other very active stars. This indicates either that the chromospheric active regions are not as optically thick as in the 
Sun or that the hemisphere averaged chromospheric emission is the result of a ``mixture'' of plage-like and prominence-like regions, the latter having a much 
higher Balmer decrement. All the few accreting stars in Cha\,I observed with UVES display a Balmer decrement of $\sim$5--30, indicating an optically thin emission 
from the accreting matter.
\item The accreting stars in Cha~I display a wide range of $r$ values, but all the stars for which we found a veiling larger than 0.25 are accretors.
\item On the one hand, the luminosity in the H$\alpha$ line proves itself to be a diagnostic more reliable than the H$\alpha$ 10\% width to derive the mass 
accretion rate, as found in previous works. On the other hand, the H$\alpha$ 10\% width represents a fast and efficient criterion to select accretor candidates 
for an ad-hoc analysis, e.g. by searching for the value of veiling that, in combination with that of other free parameters, best matches the observations. 
\end{itemize}

 Concluding, the results presented in this work, which are based on the first two young clusters observed by the GES, show the huge potential of the survey  
for the study of fundamental properties of PMS stars, such as their rotation, magnetic activity, and mass accretion properties as a function of basic stellar parameters
like mass and  age. This type of analysis can be extended to the other young clusters	that are being observed within the GES. 
This will provide an unprecedented picture of these phenomena in low-mass stars during the first stages of their evolution.
 
\begin{acknowledgements}
The authors are grateful to the referee for carefully reading the paper and for her/his useful remarks. 
This work was partly supported by the European Union FP7 programme through ERC grant number 320360 and by the Leverhulme Trust through grant RPG-2012-541. 
We acknowledge the support from INAF and Ministero dell'Istruzione, dell' Universit\`a e della Ricerca (MIUR) in the form of the grant ``Premiale VLT 2012''. 
The results presented here benefit from discussions held during the Gaia-ESO workshops and conferences supported by the ESF (European Science Foundation) 
through the GREAT Research Network Programme. 
S.G.S. acknowledge the support from the Funda\c{c}\~ao para a Ci\^encia e Tecnologia, FCT (Portugal)
in the form of the fellowship SFRH/BPD/47611/2008.
This research made also use of the SIMBAD database, operated at the CDS (Strasbourg, France) and of the Deep 
Near Infrared Survey of the Southern Sky (DENIS) database.
\end{acknowledgements}

\bibliographystyle{aa}

\Online

\topmargin 2 cm

\onecolumn

\begin{landscape}
\setlength{\tabcolsep}{3pt}


\hspace{-2cm}
\begin{longtable}{lccccccrrrrrrrrrrrrrrrrrrr}
\caption[ ]{Parameters of the $\gamma$\,Vel members used for this study.}\\
\end{longtable}

\normalsize
\addtocounter{table}{-1}

\begin{longtable}{lccccccllrrrrrrrrrrrrrrrrr}
\caption[ ]{{\it Continued.}}\\
\end{longtable}

\normalsize
\addtocounter{table}{-1}

\begin{longtable}{lccccccllrrrrrrrrrrrrrrrrr}
\caption[ ]{{\it Continued.}}\\
\end{longtable}

\normalsize 

\newpage

\begin{longtable}{lccccccllrrrrrrrrrrrrrrrrr}
\caption[ ]{Parameters of the Cha\,I members used for this study.}\\
\end{longtable}

\addtocounter{table}{-1}

\begin{longtable}{lccccccllrrrrrrrrrrrrrrrrr}
\caption[ ]{{\it Continued.}}\\
\end{longtable}


\end{landscape}

\begin{table}
\caption{Elodie templates.  }
\label{Tab:Elodie_templ}
\tiny
\begin{tabular}{llrrr}
\hline
\hline
\noalign{\smallskip}
  Name      &  {\it SpT}   & $T_{\rm eff}$ & $\log g$ &  [Fe/H]  \\  
            &         &   (K)         & \multicolumn{2}{c}{(dex)}  \\	            
\hline
\noalign{\smallskip}
  HD338529  &  F            &	6239	&  3.87  &   -2.34  \\  
  HD284248  &  F2	    &	6179	&  4.39  &   -1.60  \\  
  HD216143  &  G5	    &	4525	&  1.00  &   -2.25  \\  
  HD204543  &  G0	    &	4650	&  0.71  &   -1.86  \\  
  HD195636  &  sdG0	    &	5478	&  3.40  &   -2.79  \\  
  HD175305  &  G5III	    &	4936	&  2.08  &   -1.49  \\
\end{tabular}
\end{table}

\addtocounter{table}{-1}

\begin{table}
\caption{{\it Continued.} }
\tiny
\begin{tabular}{llrrr}
\hline
\hline
\noalign{\smallskip}
  Name      & {\it SpT}   & $T_{\rm eff}$ & $\log g$ &  [Fe/H]  \\  
            &         &   (K)         & \multicolumn{2}{c}{(dex)}  \\	            
\hline
\noalign{\smallskip}
  HD012235  &  G2IV	    &	6005	&  4.18  &    0.21  \\  
  HD013403  &  G3V	    &	5576	&  3.98  &   -0.35  \\  
  HD013783  &  G8V	    &	5493	&  4.32  &   -0.54  \\  
  HD013974  &  G0V	    &	5721	&  4.29  &   -0.44  \\  
  HD014214  &  G0.5IV	    &	6062	&  4.16  &    0.16  \\  
  HD017382  &  K1V	    &	5240	&  4.49  &   -0.04  \\
\end{tabular}
\end{table}

\addtocounter{table}{-1}

\begin{table}
\caption{{\it Continued.} }
\tiny
\begin{tabular}{llrrr}
\hline
\hline
\noalign{\smallskip}
  Name      &  {\it SpT}   & $T_{\rm eff}$ & $\log g$ &  [Fe/H]  \\  
            &         &   (K)         & \multicolumn{2}{c}{(dex)}  \\	            
\hline
\noalign{\smallskip}
  HD131977  &  K4V	    &	4395	&  4.36  &    0.07  \\  
  HD134083  &  F5V	    &	6435	&  4.19  &   -0.02  \\  
  HD142091  &  K1IV	    &	4864	&  3.40  &    0.09  \\  
  HD145675  &  K0V	    &	5261	&  4.45  &    0.35  \\  
  HD146233  &  G2V	    &	5818	&  4.45  &    0.04  \\  
  HD149661  &  K0V	    &	5315	&  4.57  &   -0.05  \\
\end{tabular}
\end{table}

\addtocounter{table}{-1}

\begin{table}
\caption{{\it Continued.} }
\tiny
\begin{tabular}{llrrr}
\hline
\hline
\noalign{\smallskip}
  Name      &  {\it SpT}   & $T_{\rm eff}$ & $\log g$ &  [Fe/H]  \\  
            &         &   (K)         & \multicolumn{2}{c}{(dex)}  \\	            
\hline
\noalign{\smallskip}
  HD132142  &  K1V	    &  5182	& 4.53   &  -0.44   \\  
  HD010700  &  G8V	    &  5290	& 4.43   &  -0.56   \\  
  HD000245  &  G2IV	    &	5722	&  3.93  &   -0.59  \\  
  HD006920  &  F9IV	    &	5922	&  3.48  &   -0.07  \\  
  HD018768  &  G1IV	    &	5769	&  3.68  &   -0.54  \\  
  HD043318  &  F6IV	    &	6196	&  3.68  &   -0.21  \\
\end{tabular}
\end{table}

\begin{table}
\caption{H$\beta$ equivalent widths and fluxes for the members of $\gamma$~Vel and Cha\,I observed with UVES.  }
\label{Tab:Hbeta}
\tiny
\begin{tabular}{lrrrr}
\hline
\hline
\noalign{\smallskip}
2MASS  &   $EW_{\rm H\beta}$ & err & $F_{\rm H\beta}$ & err \\	            
            &  \multicolumn{2}{c}{(\AA) }& \multicolumn{2}{c}{(erg\,cm$^{-2}$s$^{-1}$)}  \\	            
\hline
\noalign{\smallskip}
\multicolumn{5}{c}{$\gamma$Vel}  \\
  J08091875-4708534   &     0.046   &	0.011  &   6.058$\times 10^{5}$  &   1.574$\times 10^{5}$  \\ 
  J08092627-4731001   &     0.373   &	0.023  &   2.294$\times 10^{6}$  &   2.422$\times 10^{5}$  \\ 
  J08093304-4737066   &     0.092   &	0.018  &   8.303$\times 10^{5}$  &   1.976$\times 10^{5}$  \\ 
  J08094221-4719527   &     0.940   &	0.084  &   4.900$\times 10^{6}$  &   7.406$\times 10^{5}$  \\ 
  J08110285-4724405   &     0.185   &	0.030  &   1.019$\times 10^{6}$  &   1.915$\times 10^{5}$  \\
\end{tabular}
\end{table}
\normalsize

\begin{table}
\caption{Mass accretion rates derived in the literature with different methods.  }
\label{tab:Macc_literature}
\tiny
\begin{tabular}{cccc}
\hline
\hline
\noalign{\medskip}
2MASS  & $\log \dot M_{\rm acc}$	 & Method & Reference \\	            
       &  ($M_\odot$ yr$^{-1}$)	 &  &  \\	            
\noalign{\medskip}
\hline
\noalign{\medskip}
J10555973-7724399 & $-7.39^{+0.18}_{-0.17}$ & Br$\gamma$		      & (1) \\
 		  & $-8.8\pm0.3$	    & H$\alpha$ photometry	      & (3) \\
 		  & $-8.37\pm0.6$	    & Spectrophotometry 	      & (7)   \\
 		  & $-9.11\pm0.5$	    & $EW_{\rm H\alpha}$		      & This work \\
J10563044-7711393 & $-8.23\pm0.6$	    & Spectrophotometry 	      & (7)   \\	     
 		  & $-8.82\pm0.5$	    & $EW_{\rm H\alpha}$		      &  This work \\
\end{tabular}
~\\
Notes: (1): \cite{daemgenetal2013}; (2): \cite{costiganetal2012}; (3): \cite{robbertoetal2012}; (4): \cite{espaillatetal2011}; 
(5): \cite{antoniuccietal2011}; (6): \cite{kimetal2009}; (7): \cite{hartmannetal1998}.
\end{table}
\normalsize

\newpage
\topmargin -1 cm
\twocolumn
\begin{appendix}

\section{Cha~I: comparing $\dot M_{\rm acc}$ with the literature}
\label{sec:macc_literature}

\begin{figure}  
\begin{center}
\includegraphics[width=9cm]{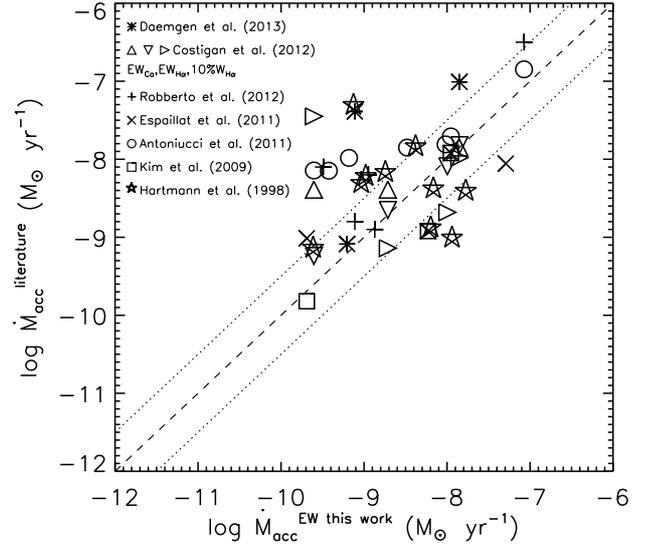}
\caption{Comparison between our $\dot M_{\rm acc}$ values  calculated using the $EW_{\rm H\alpha}$ and those obtained by several authors. Dashed and dotted lines 
represent the one-to-one relation and the position of the typical mean error in $\dot M_{\rm acc}$ of $\pm 0.5$ dex. 
The legend in the upper left corner explains the meaning of the symbols.
}
\label{fig:literature_Macc}
 \end{center}
\end{figure}

In Fig.~\ref{fig:literature_Macc}, we compare the mass accretion rates  from the literature 
with those computed in this work from the H$\alpha$ EW (see also Table
\ref{tab:Macc_literature}). 

\cite{hartmannetal1998} derived mass accretion rates from intermediate-resolution spectrophotometry of 
the hot continuum emission.  Ten accretors are in common with us. 		
Our  values and those obtained by these authors are in agreement within the errors with the only exception of 
J11072825$-$7652118, for which our $\dot M_{\rm acc}$ is lower than the \cite{hartmannetal1998} value by $\sim 1.7$ dex, but 
it is close to the values reported by other authors (see \citealt{daemgenetal2013}). Three accretors of our sample have been also 
observed by \cite{kimetal2009}, who measured $\dot M_{\rm acc}$ through $U$-band photometry.		
Differences between these values and our determinations are within $\sim 0.3$ dex, on average.
Recently, \cite{espaillatetal2011} have measured 
$\dot M_{\rm acc}$ with a similar method like the latter authors; for the four accretors in common with us a mean 
difference of $\sim 0.5$~dex is found. Seven accreting objects are in common with \cite{antoniuccietal2011}, who 
measured $\dot M_{\rm acc}$ through the Br$\gamma$ line. Four stars show similar mass accretion rates, while the values for the 
three targets with the lowest $\dot M_{\rm acc}$ are higher than ours. Similar differences have been found also 
by \cite{biazzoetal2012} for low-mass stars in Chamaeleon~II. \cite{robbertoetal2012} have derived $\dot M_{\rm acc}$ 
from H$\alpha$ photometry for five accretors of our sample. The mean difference in $\dot M_{\rm acc}$ between 
their and our value is $\sim 0.7$~dex.  \cite{costiganetal2012} report $\dot M_{\rm acc}$ measurements 
using three different diagnostics ($EW_{\rm H\alpha}$, $10\%W_{\rm H\alpha}$, and $EW_{\rm \ion{Ca}{ii}-\lambda8662}$) for 
four accretors in common with us. The agreement between their results and ours is good, especially when we consider the 
$\dot M_{\rm acc}$ derived from their $EW_{\rm H\alpha}$. 
 The case of J11075809$-$7742413 is emblematic because they measure the highest difference in $\dot M_{\rm acc}$ derived 
through the three methods, but the value obtained with $EW_{\rm H\alpha}$ is very close to our one. 
This suggests that the discrepancies in the $\dot M_{\rm acc}$ values are mostly due to the method used for deriving
it rather than to the different instrumentation used or to an intrinsical variability of the source.
Finally, \cite{daemgenetal2013} have observed three accretors in common with us and have adopted the Br$\gamma$ line as  
diagnostics. The agreement with our values is fairly good, with the exception of 10555973$-$7724399 
for which they have derived $\log \dot M_{\rm acc}=-7.4 M_\odot$\,yr$^{-1}$ at odds with our value of $-9.1 M_\odot$\,yr$^{-1}$, 
which is more similar, within the errors, to the values of $-8.8 M_\odot$\,yr$^{-1}$ and $-8.4 M_\odot$\,yr$^{-1}$ obtained 
by \cite{robbertoetal2012} and \cite{hartmannetal1998}, respectively.

Concluding, we think that the comparison between our $\dot M_{\rm acc}$, as derived from the H$\alpha$ luminosity, and the literature values 
is in general quite good.  The differences/inconsistencies can be attributed to intrinsic short-term and long-term 
variability (as outlined in Sect.~\ref{sec:macc_mass}) and the different photometric/spectroscopic methodologies used by each 
author to derive accretion luminosity and mass accretion rate, as well as to the different evolutionary models adopted 
to estimate the stellar parameters (as also recently pointed out by \citealt{alcalaetal2014}). 

\end{appendix}

\end{document}